\documentclass[%
 reprint,
 amsmath,amssymb,
 aps,
prl,
nofootinbib,
prstper,
floatfix,
]{revtex4-1}

\usepackage{graphicx}
\usepackage{dcolumn}
\usepackage{bm}
\usepackage{color}
\usepackage[colorlinks]{hyperref}
\usepackage{dsfont}
\usepackage{pst-node}%
\usepackage{array,mathtools,amssymb,booktabs}
\usepackage[utf8]{inputenc}
\usepackage[T1]{fontenc}
\begin{document}

\title[]{Effect of Resource Dynamics on Species Packing in Diverse Ecosystems}

\author{Wenping Cui}
\email{cuiw@bu.edu}
\affiliation{Department of Physics, Boston University, 590 Commonwealth Avenue, Boston, MA 02139}
\affiliation{ Department of Physics, Boston College, 140 Commonwealth Ave, Chestnut Hill, MA 02467}

\author{Robert Marsland III}
\email{marsland@bu.edu}
\affiliation{Department of Physics, Boston University, 590 Commonwealth Avenue, Boston, MA 02139}

\author{Pankaj Mehta}%
\email{pankajm@bu.edu}
\affiliation{Department of Physics, Boston University, 590 Commonwealth Avenue, Boston, MA 02139}
\date{\today}
       

\begin{abstract}
The competitive exclusion principle asserts that coexisting species must occupy distinct ecological niches (i.e. the number of surviving species can not exceed the number of resources). An open question is to understand if and how different resource dynamics affect this bound. Here, we analyze a generalized consumer resource model with externally supplied resources and show that -- in contrast to self-renewing resources -- species can occupy only half of all available environmental niches. This motivates us to construct a new schema for classifying ecosystems based on species packing properties.
\end{abstract}

\maketitle
\newpage
\clearpage


One of the most stunning aspects of the natural world is the incredible diversity of species present in many environments \cite{huttenhower2012structure, gentry1988tree}. A major goal of community ecology is to understand the rules governing community structure and species coexistence patterns in these complex ecosystems. One promising approach that has recently emerged for tackling this challenge is to use ideas from statistical mechanics  inspired by spin glass physics 
 \cite{mezard1987spin, nishimori2001statistical}. In such an approach, ecosystems are viewed as large interacting disordered systems, allowing for the identification of universal, collective properties  \cite{barbier2018generic, cui2019diverse}.  Such statistical physics inspired models  are also able to reproduce many  experimental observations, especially in the context of microbial ecosystems \cite{goldford2018emergent,
 marsland2019available, marsland2019minimal}.

Much of this work has focused on generalized Lotka-Volterra models where species directly interact with each other in a pair-wise fashion \cite{fisher2014transition,kessler2015generalized, bunin2017ecological, kessler2015generalized, bunin2017ecological, Barbier147728, barbier2018generic, biroli2018marginally, roy2019numerical, pearce2019stabilization}.   While such models have led to deep ecological insights \cite{chesson2000mechanisms} and have allowed for the identification of interesting ecological phases 
and phase transitions \cite{fisher2014transition,kessler2015generalized, bunin2017ecological}, a major drawback of Lotka-Volterra models are that they do not explicitly model the resources present in the ecosystem. Instead, resource dynamics are implicitly represented through the choice of species-species interactions making it difficult to understand the relationship between resource dynamics and community structure. 
\begin{figure}
\centering
\includegraphics[width=0.45\textwidth]{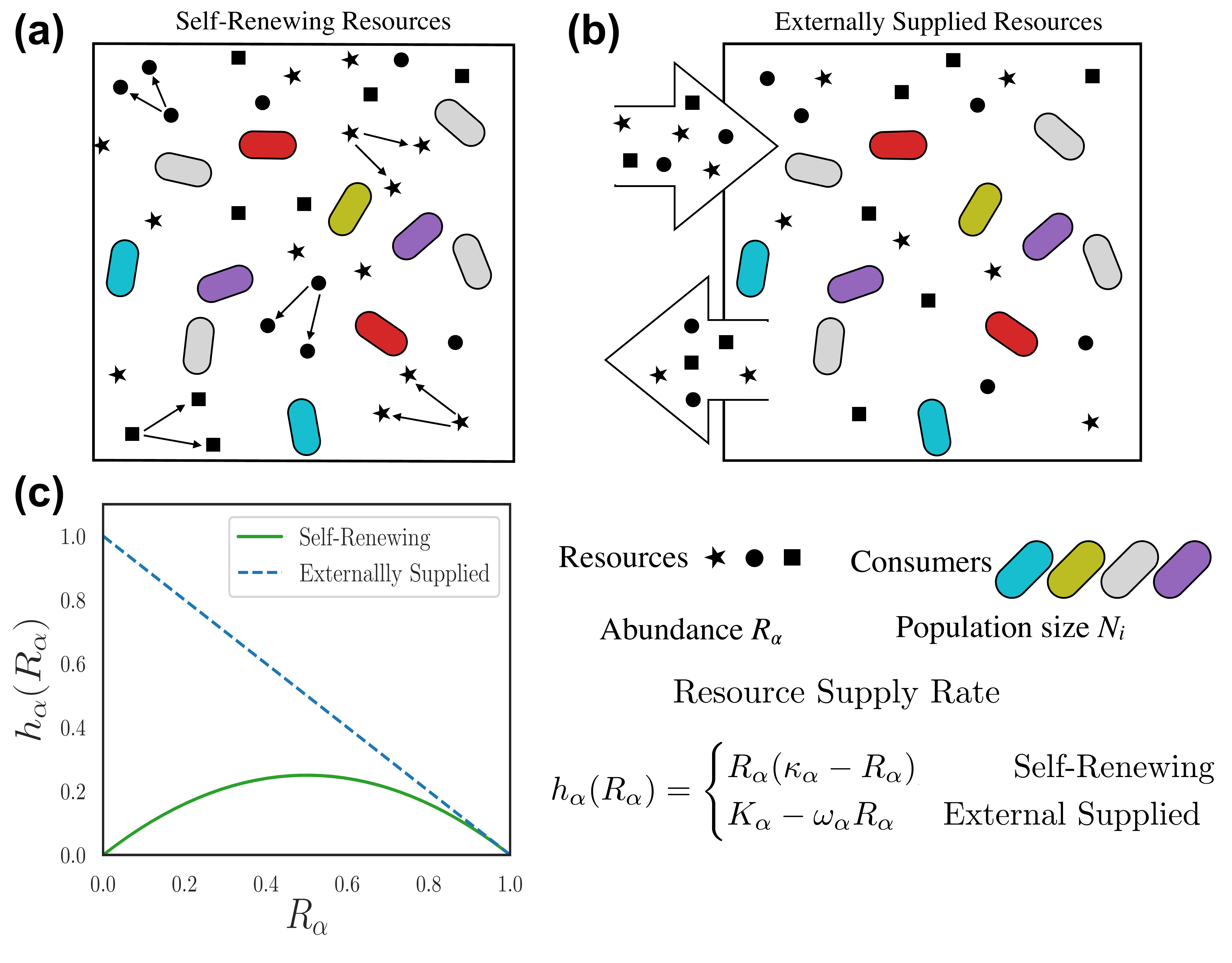}
\caption{Schematic description for two types of resources. (a) Self-renewing resources (e.g. plants), which are replenished through organic reproduction;
(b) Externally supplied resources (e.g. nutrients that sustain gut microbiota), which are replenished by a constant flux from some external source, and diluted at a constant rate;  (c) The supply rate as a function of resource abundance for both choices, with $\kappa=\omega_\alpha=K_\alpha=1$.}
\label{scheme}
\end{figure}

In contrast, generalized consumer-resource models (GCRMs), first introduced by MacArthur and Levins in a series of seminal papers \cite{macarthur1967limiting, chesson1990macarthur, macarthur1970species}, explicitly incorporate both species and
and resource dynamics. In GCRMs, ecosystems are described by species that can consume and deplete resources according to a set of consumer preferences. Interactions between species arise because species with similar consumer 
preferences occupy similar environmental niches and hence compete for common resources. An important theoretical and conceptual result that follows from GCRMs  is that the number of species that can coexist in an ecosystem is limited by the number of resources that are present. In other words, if we denote the number of species that can survive in an ecosystem by  $S^*$ and  the number of supplied resources as $M$, the competitive exclusion principle yields
an upper bound for the amount of species that can be packed into the ecosystem: $\frac{S^*}{M} \le1$\cite{mcgehee1977some}. 

The basic intuition behind this bound is that the growth rates $g_i(\mathbf{R})$ of all coexisting species $i=1,2,\dots$ must simultaneously vanish, and since the space of resource concentrations $\mathbf{R}$ is $M$-dimensional, at most $M$ of these equations can be simultaneously solved (see Supplemental Material(SM) for discussion of non-generic phenomena where the bound is violated).
While this result gives {an upper bound},  it is not clear when and if it will be saturated. In particular, we show below that the choice of resource dynamics  fundamentally alters
species-packing properties. To show this, we analyze GCRMs with two different resource dynamics:  \emph{self-renewing} resources where resources grow logistically in the absence of consumers \cite{macarthur1967limiting, chesson1990macarthur} and \emph{externally supplied} resources that are supplied and degraded at a constant rate \cite{posfai2017metabolic,  tikhonov2017collective, marsland2019minimum} (see Fig. \ref{scheme}).  We derive species packing bounds for both choices of dynamics by analyzing the susceptibilities of a new cavity solution for GCRMs with externally supplied resources and combining it with the previously derived cavity solution for GCRMs with self-renewing resources \cite{advani2018statistical, mehta2019constrained,cui2019diverse}.  Surprisingly, in the absence of metabolic tradeoffs we find that, for externally-supplied resources, species can occupy only half of all available resource niches:  $\frac{S^*}{M}<\frac{1}{2}$. Motivated by these results, we suggest a new schema for classifying ecosystems based on their species packing properties.


\noindent {\bf Model:} GCRMs describe the ecological dynamics of $S$ species of consumers $N_i$ ($i=1, 2, \ldots S$) that can consume $M$ distinct resources $R_\alpha$ ($\alpha=1, 2, \dots , M$). The
 rate at which species $N_i$ consumes and depletes resource $R_\beta$ is encoded in a matrix of consumer preferences $C_{ i \beta}$.  In order to survive, species have a minimum
 maintenance cost $m_i$. Equivalently, $m_i$ can also be thought of as the death rate of species $i$ in the absence of resources.  These dynamics can be described using a coupled set 
 of $M+S$ ordinary differential equations of the form
\begin{eqnarray}
\begin{cases}
&\frac{d{N}_{i}}{dt}  =N_{i}\sum_{\beta} C_{i \beta} R_{\beta} - N_i m_i\\
&\\
&\frac{d{R}_{\alpha}}{dt}  =h_\alpha(R_\alpha)- \sum_{j}N_{j} C_{j\alpha}R_{\alpha} ,
\end{cases}\label{CRM0} 
\end{eqnarray} 
where $h_\alpha(R_\alpha)$ a function that describes the dynamics of the resources in the absence of any consumers (see Fig. \ref{scheme}).

For  self-renewing resources (e.g. plants, animals), the dynamics can be described using logistic growth of the form
 \begin{equation}
 h_\alpha(R_\alpha)=R_\alpha(\kappa_{\alpha} -  R_\alpha),
 \label{eq:logistic}
 \end{equation}
with $\kappa$  the carrying capacity. While such resource dynamics is reasonable for biotic resources, abiotic resources such as minerals and small molecules cannot self-replicate and are usually supplied externally to the ecosystem ( Fig. \ref{scheme}(b)). A common 
way to model this scenario is by using linearized  resource dynamics of the form
\begin{equation}
h_\alpha(R_\alpha)=K_{\alpha} - \omega_\alpha R_\alpha.
 \label{eq:linear}
\end{equation}
Fig. \ref{scheme}(c) shows a plot of these two choices. Notice that the two resource dynamics behave very differently at low resource levels. The self-renewing resources can go extinct and eventually disappear from the ecosystem while this is not true of 
externally supplied resources.

Recent research has shown some unexpected and interesting non-generic
phenomena can appear in GCRMs  in the presence of additional constraints on parameter values. A common choice of such constraints is the  imposition of a ``metabolic budget'' on the consumer preference matrix \cite{posfai2017metabolic,  li2019modeling}  tying the maintenance cost $m_i$ to the total consumption capacity $\sum_\beta C_{i\beta}$  \cite{tikhonov2017collective, altieri2019constraint}. These metabolic tradeoffs can be readily incorporated into the cavity calculations and have significant impacts on species packing as will be discussed below.

\noindent {\bf Cavity solution:} Recently, we derived a mean-field cavity solution for steady-state dynamics of the the GCRM with self-renewing resource dynamics in the high-dimensional limit where the number of resources and species in the regional species pool is large ($S, M  \gg 1$)\cite{advani2018statistical, mehta2019constrained,cui2019diverse}. The overall procedure for deriving the cavity equations for GCRM with externally supplied resource is similar to that for GCRMs with self-renewing resources and is shown in Fig. S1 in the SM. We assume the $K_\alpha$ and $m_i$ are independent random normal variables with means $K$ and $m$ and variances $\sigma^2_K$  and $\sigma^2_m$, respectively. We also assume $\omega_\alpha$ are independent normal variables with mean $\omega$ and variance $\sigma_\omega^2$. The elements of the consumption matrix $C_{i\alpha}$ are drawn independently from a normal distribution with mean $\mu/M$ and variance $\sigma^2_c/M$.  
This scaling with $M$ is necessary to guarantee that  $\left< N\right>$, $\left< R\right>$ do not vanish when $S, M  \gg 1$ with $M/S = \gamma$ fixed. Later, we will consider a slightly modified scenario where the maintenance costs are correlated with the consumption matrix in order to implement the metabolic tradeoffs discussed above.

The basic idea behind the cavity method is to derive self-consistency equations relating an ecosystem with $M$ resources and $S$ species to an ecosystem with $M+1$ resources and $S+1$ resources. This is done by adding  a new "cavity"  species 0 and a new "cavity"  resource 0 to the original ecosystem. When $S,M \gg 1$, the effect of the new cavity species/resource is small and can be treated using perturbation theory.  The cavity solution further exploits the fact
that since the $C_{i \alpha}$ are random variables, when $M \gg 1$ the sum  $\sum_\alpha C_{i\alpha}R_\alpha$ will be well described by a by a normal distribution with mean $\mu\left< R\right>$ and variance $\sigma_c^2 q_R$ where $q_R=\left<R^2\right>= 1/M \sum_\alpha R_\alpha^2$ (see SM for details). Combining this with the non-negativity constraint, the species distribution can be expressed as a truncated normal distribution, 
\begin{eqnarray}\label{eq:N}
\bar{N}=\mathrm{max}\left[ 0, \frac{\mu\left< R\right>-m+\sqrt{\sigma_c^2 q_R+\sigma_m^2}z_N}{\sigma_c^2 \chi}\right]
\end{eqnarray}
where $\chi=-\left< \frac{\partial \bar{R}_\alpha}{\partial \omega_\alpha}\right> =-M^{-1} \sum_\alpha \frac{\partial \bar{R}_\alpha}{\partial \omega_\alpha} $ and $z_N$ is a standard normal variable. This equation describes GCRMs with both externally supplied and self-renewing resource dynamics \cite{advani2018statistical}.

The steady-state cavity equations for externally supplied resources are significantly more complicated and technically difficult to work with than the corresponding equations for self-renewing resources. To see this, notice  that the steady-state 
abundance of resource $\alpha$ can be found by plugging in  Eq. \ref{eq:linear} into Eq \ref{CRM0} and setting the left hand side to zero to get
\begin{equation}
\bar{R}_\alpha=K_\alpha/(\omega_\alpha+\sum_j \bar{N}_j C_{j\alpha})=  {K_\alpha \over \omega_\alpha^\mathrm{eff}},
\label{eq:ss-ex-sup}
\end{equation}
where we have defined $\omega_\alpha^\mathrm{eff}= \omega_\alpha+\sum_j \bar{N}_j C_{j\alpha}$.
When $S \gg 1$, both the denominator $\omega_\alpha^\mathrm{eff}$ and the numerator $K_\alpha$ can be modeled by independent normal random variables. This implies that the
the steady-state resource abundance is described by a ratio of normal variables (i.e. the Normal Ratio Distribution) instead of a truncated Gaussian as in the self-renewing case \cite{marsaglia2006ratios}(see Fig. S5). At large $\sigma_c$, this makes solving the cavity
equations analytically intractable.  Luckily, if the variance of the denominator $\omega_\alpha^\mathrm{eff}$ is small compared with the mean -- which is true when $\sigma_c$ not too large -- we can still obtain an approximate replica-symmetric solution 
 by expanding in powers of the standard deviation over the mean of $\omega_\alpha^\mathrm{eff}$ (see SM).  We consider expansions to the cavity solutions where the denominator in Eq. \ref{eq:ss-ex-sup} is expanded to $1^{st}$ order. In general, the backreaction correction is quite involved since resources and species form loopy interactions resulting in non-trivial correlation between $C_{i\alpha}$ and $N_i$ that must be properly accounted for (see SM).
\begin{figure}
\centering
\includegraphics[width=0.48\textwidth]{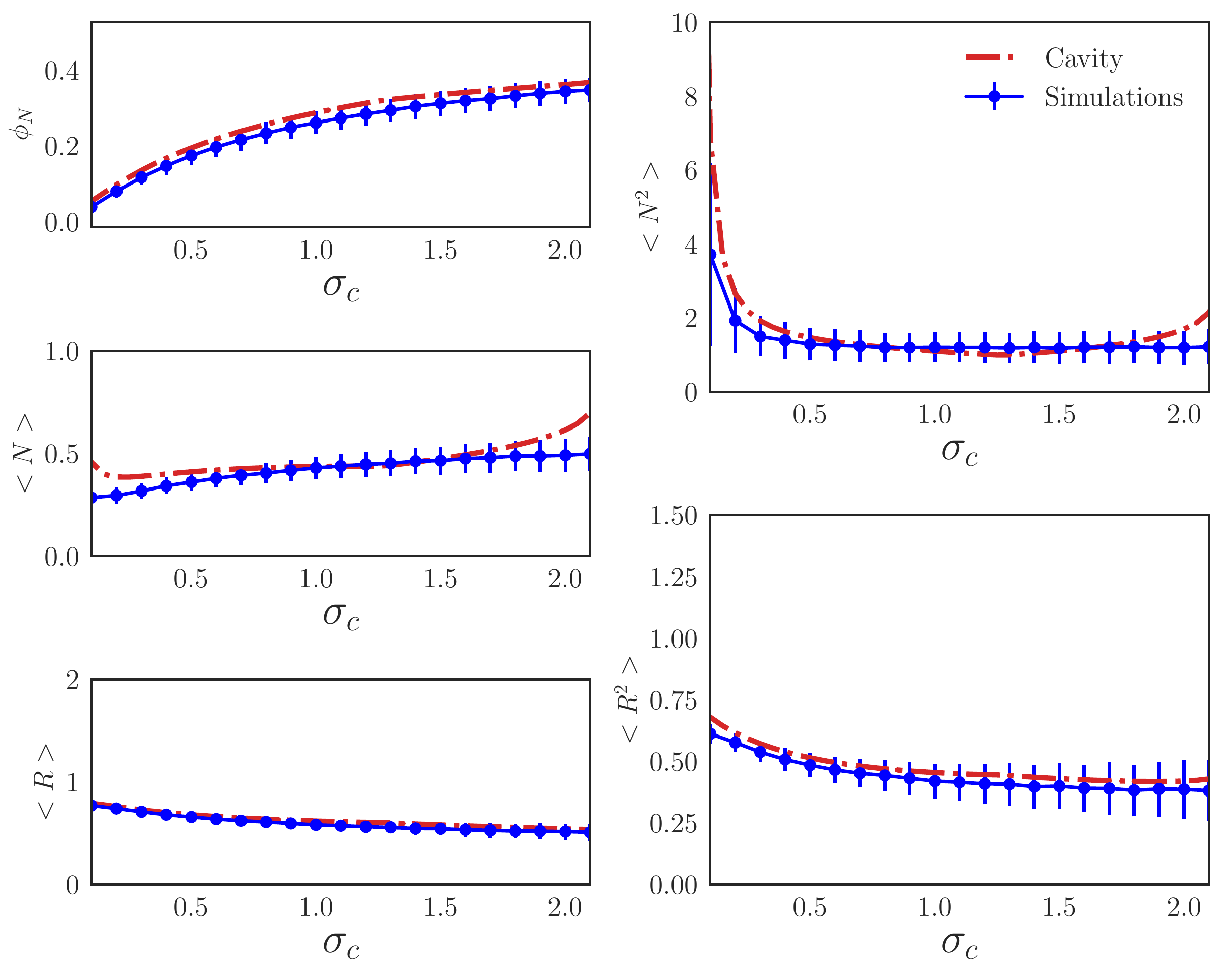}
\caption{Comparison between  cavity solutions (see main text for definition) and  simulations for the fraction of surviving species $\phi_N= \frac{S^*}{S}$
and the first and second moments of the species and resources distributions as a function of $\sigma_c$. The error bar shows the standard deviation from $1000$ numerical simulations with $M=S=100$ and all other parameters are defined in the SM. Simulations were run using the CVXPY package \cite{cvxpy_rewriting}.}
\label{solution}
\end{figure}

\noindent {\bf Comparison with numerics:} The full derivation of  $1^{st}$  order expansions of the mean-field equations are given in the SM. The resulting self-consistency equations can be solved numerically in Mathematica.  Fig. \ref{solution}
shows a comparison between the cavity solution and 1000 independent numerical simulations for various ecosystem properties such as the fraction of surviving species $S^*/S$ and the first and second moment of the species
and resource distributions (simulation details are in the SM).  As can be seen in the figure,  our analytic expressions agree remarkably well over a large range of $\sigma_c$.
However,  at very large $\sigma_c$ (not shown), the cavity solutions start deviating from the numerical simulations because 
 the Ratio Normal Distribution can no longer be described using the  $1^{st}$  order expansion to the full cavity equations. 
 
 As a further check on our analytic solution, we ran simulations where the $C_{i \alpha}$ were drawn from different distributions. One pathology of choosing $C_{i \alpha}$ from a  Gaussian distribution is that when $\sigma_c$ is large,  many of 
 consumption coefficients are negative. To test whether our cavity solution still describes ecosystems when $C_{i \alpha}$ are strictly positive,  we compare our cavity solution to simulations where the $C_{i \alpha}$  are drawn from a Bernoulli or uniform distribution.
 As before, there is remarkable agreement between analytics and numerics (see Fig. S2)

\noindent{\bf Species packing  without metabolic tradeoffs:} The essential ingredients needed to derive species packing bounds for GCRMS are the cavity equations for the average local  susceptibilities $\nu =\left<{\partial \bar{N}_i  \over \partial m_i }\right>=S^{-1} \sum_j {\partial \bar{N}_i  \over \partial m_i }$ and $\chi= \left< {\partial \bar{R}_\alpha  \over \partial X_\alpha } \right> =M^{-1} {\partial \bar{R}_\alpha  \over \partial X_\alpha }$, with $X_\alpha=K_\alpha$ for externally supplied resources and $X_\alpha=-\omega_\alpha$ for self-renewing resources.  These two susceptibilities measure how the mean  species abundance and mean resource abundance respond to changes in the species death  rate and the resource supply/depletion rate, respectively. They  play an essential role in the cavity
equation and can be used for distinguishing different phases in complex systems\cite{ramezanali2015cavity, cui2019diverse}.

For the self-renewing case, the susceptibilities $\chi_s$ and $\nu_s$ are given by eq. (59, 60) in \cite{mehta2019constrained} 
\begin{eqnarray}
\nu_s = -\frac{\phi_N}{\sigma_c^2 \chi_s}, \quad \chi_s=\frac{\phi_R}{1-\gamma^{-1}\sigma_c^2\nu_s},
\end{eqnarray}
and can be reduced to $\chi_s=\phi_R-\gamma^{-1}\phi_N$, where $\phi_R=M^*/M$, with $M^*$ equal to the number of non-extinct resources in the ecosystem. In order to guarantee the positivity of $\left< N\right>$, we must have $\chi_s=\phi_R-\gamma^{-1} \phi_N>0$, resulting in an upper bound
\begin{eqnarray}
1 \ge \frac{M^*}{M}> \frac{S^*}{M} \label{eq:bound1}
\end{eqnarray}
which states that the number of surviving resources must be larger than the number of surviving species. 

\begin{figure}
\centering
\includegraphics[width=0.5\textwidth]{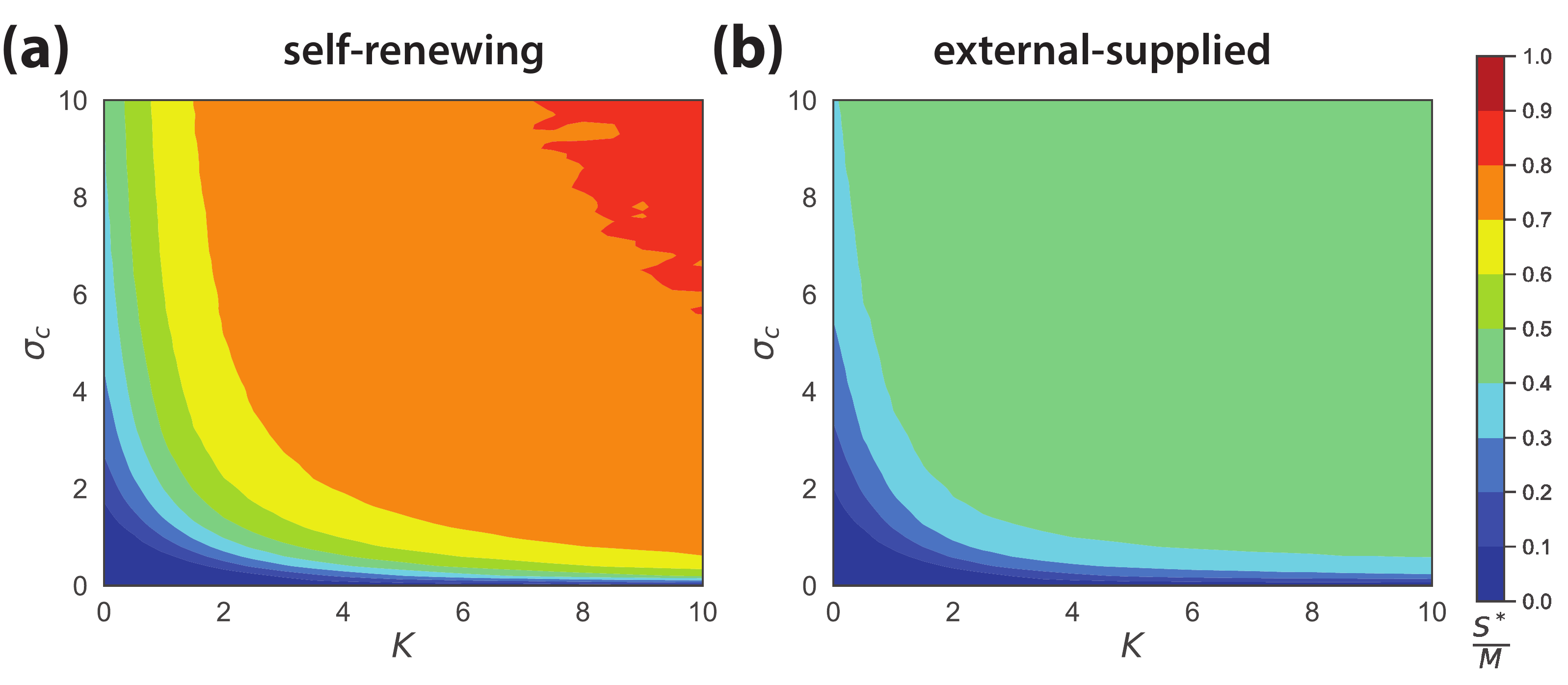}
\caption{Comparison of the species packing ratio$\frac{S^*}{M}$ at various $\sigma_c$ and $K$ for self-renewing and externally supplied resource dynamics. The simulations represent averages from 1000 independent realizations with the system size $M=100$, $S=500$ (parameters in SM). }
\label{bound1}
\end{figure}

For the externally supplied case, the corresponding equations take the form
\begin{eqnarray}
\nu&=& -\frac{\phi_N}{\sigma_c^2 \chi}, \chi= -\frac{1}{2\gamma^{-1} \nu \sigma_c^2}(1-\left<... \right>) \label{eq:chimain},
\end{eqnarray}
where  the full expression of $\left<... \right>$ can be found in eq. (63) in the SM. For our purposes, the most important 
property is that in the absence of metabolic tradeoffs, the expression $\left<... \right>$ is always \emph{positive}. Combining this observation with the equations above gives the upper bound 
\begin{equation}
\frac{1}{2} >\frac{S^*}{M}=\phi_N \gamma^{-1}.
\end{equation}
Thus, for externally supplied resources, at most \emph{half of all potential niches} are occupied.  Fig. \ref{bound1}
shows numerical simulations confirming the species packing bound for various choices of $K$ and $\sigma_c$ (see Fig. S6 in SM for various choices of $S/M$). The lower diversity found when resources are supplied externally can be anticipated by noting that the resource abundance in this model is more narrowly distributed than in a model with self-renewing resources. As a result, species experience stronger competition (see Fig. S5 and more details in SM). However, we still currently lack an intuitive explanation of why the species packing bound is exactly $0.5$.

\begin{figure}[t]
\centering
\includegraphics[width=0.5\textwidth]{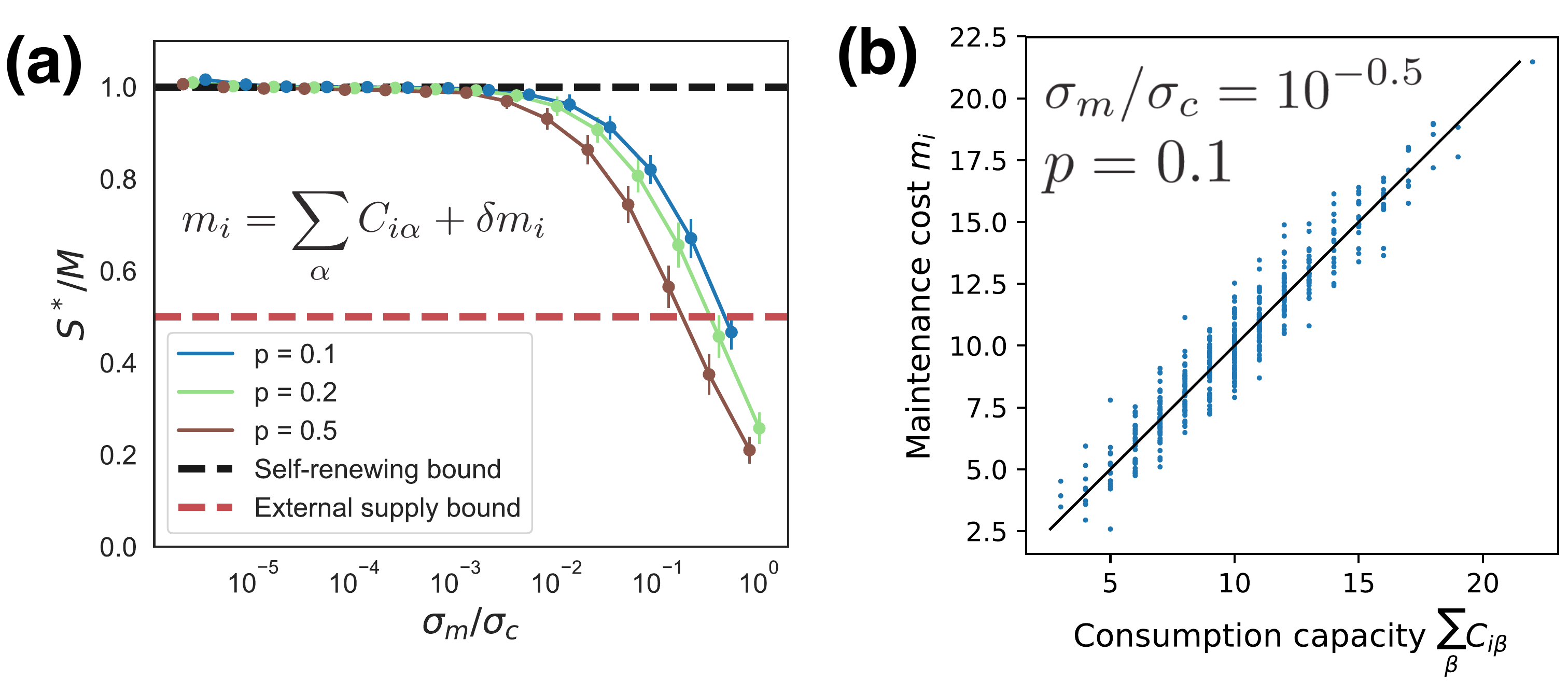}
\caption{Species packing bounds in the presence of metabolic tradeoffs. (a) The species packing ratio $S^*/M$  as a function of $\sigma_m/\sigma_c$, where $\sigma_m$ is the standard deviation of the $\delta  m_i$  and  $\sigma_c/\sqrt{M}$ is the standard deviation of
 $C_{i \alpha}$. Simulations are for binary consumer preference matrix $C_{i \alpha}$ drawn from a Bernoulli distribution with probability $p$. (b) $m_i$ versus $\sum_{\alpha}C_{i \alpha}$ for $p=0.1$ and $\sigma_m/\sigma_c=10^{-0.5}$ See SM for all parameters. }
\label{metabolic_bound}
\end{figure}

\noindent{\bf Species packing with metabolic tradeoffs:} We also find that metabolic tradeoffs modify the cavity equations in such a way that the expression in brackets $\left<... \right>$ in Equation (\ref{eq:chimain}) can become negative (see SM). However, it still remains greater than -1, allowing us to derive
a species packing bound of the form  $S^*<M$ even in the presence of soft metabolic constraints. In Figure \ref{metabolic_bound}, we simulated various ecosystems where the maintenance costs of species were chosen to obey metabolic tradeoffs of the form 
$m_i =\sum_{\alpha} C_{i \alpha} + \delta m_i$, where $\delta m_i$ are independent and identically distributed (i.i.d.) normal variables with variance $\sigma_m^2$. Note that a larger $\sigma_m$ corresponds to ecosystems with softer metabolic constraints. We found that when $\sigma_m /\sigma_c>1$, these ecosystems obey
the 1/2 species packing bound derived above. This can also be analytically shown using the modified cavity equations derived in the SM. Finally, we show in the SM that when the metabolic tradeoffs take the form of hard  constraints on the consumer preferences as in  \cite{altieri2019constraint,tikhonov2017collective,posfai2017metabolic,li2019modeling}, the cavity equations allow for interesting non-generic behavior with $S^* \geq M$, consistent with these previous works. Importantly, we find that even 
modest modifications of the tradeoff equation $m_i  \propto \sum_\alpha C_{i \alpha}$ results in ecosystems that satisfy the 1/2 species packing bound.
 
 
 \noindent{\bf Classifying ecosystems using species packing:}  Recently, it has become clear that there is a deep relationship between ecosystem and constraint satisfaction problems \cite{mehta2019constrained, marsland2019minimum, tikhonov2017collective, altieri2019constraint}. In particular, each species can be thought of as a constraint on possible resource abundances \cite{mehta2019constrained, marsland2019minimum}. Inspired by jamming \cite{liu2010jamming} , this suggests that we can
 separate ecosystems into qualitatively distinct classes depending on whether the competitive exclusion bound is saturated.  We designate ecosystems where $S^* \rightarrow M$ (like GCRMs with self-renewing resources) as
 \emph{isostatic species packings}, and ecosystems where the upper bound $S_{\rm max}$ on the number of surviving species is strictly less than the number of resources $S^*< S_{\rm max} < M$ (like GCRMs with externally supplied resources without metabolic
 tradeoffs) as \emph{hypostatic species packings}. Ecosystems with $S^*\geq M$ (like GCRMs with hard metabolic constraints) are designated as \emph{non-generic species packings} because of the presence of a macroscopic number of additional hard constraints (i.e. the number of additional constraints that are imposed scales with $S$  and $M$  in the limit $S,M \rightarrow \infty$). This basic schema suggests a way of refining the competitive exclusion principle and may help shed light on controversies surrounding the validity of basic species packing bounds.

\noindent {\bf Discussion:} In this paper, we examine the effect of resource dynamics on community structure and large-scale ecosystem level properties. To do so, we analyzed generalized Consumer Resource Models (GCRMs) with two different resource dynamics: externally supplied resources that are supplied and degraded at a constant rate and self-replicating resources whose behavior in the absence of consumers is well described
by a logistic growth law. Using a new cavity solution for GCRMs with externally supplied resources and a previously found cavity solution of the GCRM with self-renewing resources, we show that the community structure
is fundamentally altered by the choice of resource dynamics. In particular, for externally supplied resources, we find that species generically can only occupy \emph{half} of all available niches whereas for self-renewing
resources all environmental niches can be filled. We confirm this surprising bound using numerical simulations. 

In this manuscript, we consider the effect of metabolic trade-offs and show that they can increase species packing in an ecosystem. In the future, it will be interesting to ask how other specialized network structures, including niche partitioning, higher specialization, or combinations of specialists and generalists can affect our results. Based on our experience, we expect that, even in these more complicated ecosystems our species packing bound will hold quite generically. But much more work needs to be done to confirm if this is really the case.

Our results show how resource dynamics, which are neglected in commonly used Lotka-Volterra models, can fundamentally alter the properties of ecosystems. Much work still needs to be done to see if and how our results must be modified to account for other ecological processes such as demographic stochasticity, spatial structure, and microbe-specific interactions such as cross-feeding \cite{goldford2018emergent, marsland2019available}. It will also be necessary to move beyond steady-states and consider the dynamical properties of these ecosystems. More generally, it will be interesting to further explore the idea that we can classify ecosystems based on species-packing properties and see if such a schema can help us better understand the origins of the incredible diversity we observe in real-world ecosystems.

\noindent {\bf Acknowledgments:} The work was supported by NIH NIGMS grant 1R35GM119461, Simons Investigator in the Mathematical
Modeling of Living Systems (MMLS). The authors also acknowledge support from the SSC computing cluster at BU.

\bibliography{ref.bib}

\pagebreak
\widetext
\begin{center}
\textbf{\large Supplemental Materials}
\end{center}

\appendix
\onecolumngrid
\renewcommand{\thefigure}{S\arabic{figure}}

\section{Derivation of cavity solution}\label{Cavity}
\begin{figure}[h!]
\centering
\includegraphics[width=1.\textwidth]{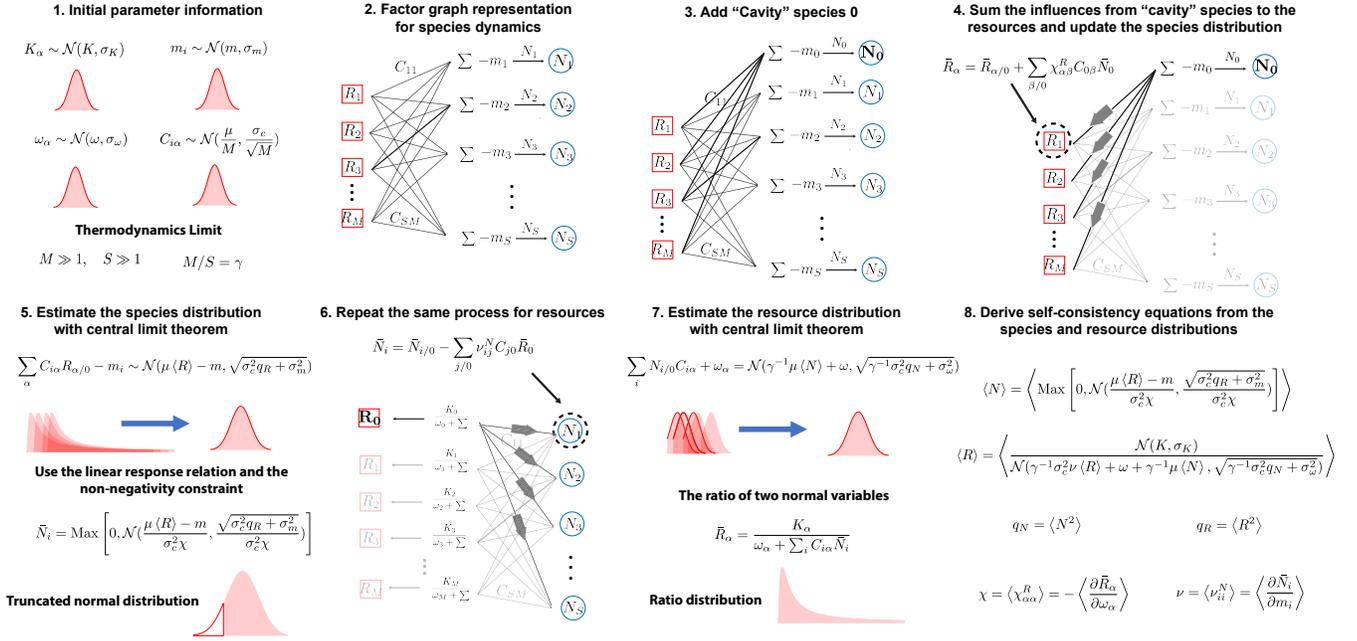}
\caption{Schematic outlining steps in cavity solution. \textbf{1.} The initial parameter information consists of the probability distributions for the mechanistic parameters: $K_\alpha$, $m_i$ and $C_{i\alpha}$. We assume they can be described by their first and second moments.   \textbf{2.} The species dynamics $N_i(\sum_{\alpha}C_{i\alpha}R_\alpha-m_i)$ in eqs. (\ref{CRM}) are expressed as a factor graph.  \textbf{3.} Add  the "Cavity" species $0$ as the perturbation. \textbf{4.} Sum the resource abundance perturbations from the "Cavity" species 0 at  steady state and update the species abundance distribution to reflect the new steady state. \textbf{5.}  Employing the central limit theorem, the backreaction contribution from the "cavity" species 0  and the non-negativity constraint, the species distribution is expressed as a truncated normal distribution. \textbf{6.} Repeat \textbf{Step 2-4} for the resources.   \textbf{7.} The resource distribution is the ratio distribution from the ratio of two normal variables $K_\alpha$ and $\omega_\alpha+\sum_i N_i C_{i\alpha}$.  \textbf{8.} The self-consistency equations are obtained from the species and resource distributions. Note that $\gamma^{-1}\sigma_c^2\nu\left< R \right>$ in the dominator of $\left<R \right>$ is from the correlation between $N_i$ and $C_{i\alpha}$ in $\sum_i N_i C_{i\alpha}$. }
\label{cavity}
\end{figure}
\subsection{Model setup}
In this section, we derive the cavity solution to the linear resource dynamics (eq. 1)  in the main text)
\begin{eqnarray}
\begin{cases}
&\frac{d{N}_{i}}{dt}  =N_{i}\left(\sum_{\beta} C_{i \beta} R_{\beta} - m_i\right)\\
&\\
&\frac{d{R}_{\alpha}}{dt}  =K_{\alpha} - \omega_\alpha R_\alpha- \sum_{j}N_{j} C_{j\alpha}R_{\alpha}  
\end{cases}\label{CRM} 
\end{eqnarray} 
Note that here we follow closely our derivation in \cite{advani2018statistical,mehta2019constrained}. The main difference is that here we consider  linear resource dynamics, which as we will see below,
makes the problem much more technically challenging.

Consumer preference $C_{i\alpha}$ are random variables drawn from a Gaussian distribution with mean $\mu/M$ and variance $\sigma_c^2/M$.
They can be deposed into $C_{i\alpha}=\mu/M + \sigma_c d_{i\alpha}$, where the fluctuating part $d_{i\alpha}$ obeys
\begin{eqnarray} \label{eq:d-mean}
\left< d_{i\alpha} \right> &=&0\\ \label{eq:d-var}
\left< d_{i\alpha} d_{j\beta} \right> &=&\frac{\delta_{ij}\delta_{\alpha\beta}}{M}.
\end{eqnarray}
We also assume that both the carrying capacity $K_\alpha$ and the minimum maintenance cost $m_i$ are independent Gaussian random variables with mean and covariance given by 
\begin{eqnarray}
\left< K_\alpha\right> &=& K\\
\text{Cov} (K_\alpha, K_\beta) &=& \delta_{\alpha\beta}\sigma^2_K\\
\left< m_i\right> &=& m\\
\text{Cov} (m_i, m_j) &=& \delta_{ij}\sigma^2_m
\end{eqnarray}

Let $\left< R\right>=\frac{1}{M}\sum_\beta R_\beta$ and $\left< N\right>=\frac{1}{S}\sum_j N_j$ be the average resource and average species abundance, respectively. With all these defined, we can re-write eqs. (\ref{CRM}) as

\begin{eqnarray} 
\frac{d N_i}{  dt} &=& N_i \left\{ \mu\left< R\right>-m+ \sum_{\beta}\sigma_c d_{i \beta} R_{\beta}-\delta m_i\right\}\label{eq-r}\\ 
\frac{d R_\alpha}{dt}\!&=& K + \delta K_\alpha - \left[\omega_\alpha+ \gamma^{-1}\mu\left< N\right>+\sum_j\sigma_c d_{j\alpha}N_j\right]R_\alpha\label{eq-n} 
\end{eqnarray} 
where $\delta K_\alpha = K_\alpha  -K, \delta m_i = m_i-m$ and $\gamma = M/S$. As noted in the main text, the basic idea of cavity method is to relate an ecosystem with $M+1$ resources (variables) and $S+1$ species (inequality constraints) to that with $M$ resources and $S$ species. Following eq. (\ref{eq-r}) and eq. (\ref{eq-n}), one can write down the ecological model for the $(M+1, S+1)$ system where resource $R_0$ and species $N_0$ are introduced to the $(M,S)$ system as: 
\begin{eqnarray} 
\frac{d N_0}{  dt} \!&=& \!N_0\left\{\mu\left< R\right>\!-\!m\!+\! \sum_{\beta}\sigma_c d_{0\beta} R_{\beta}\!-\!\delta m_0\right\}\label{dy-N0} \\
\frac{d R_0}{dt}\!&=& \!K \!+\! \delta K_0 -\left[\omega_0+ \gamma^{-1}\mu\left< N\right> +\sum_j \sigma_c d_{j0}N_j \right]R_0  \label{dy-R0}
\end{eqnarray} 
\subsection{Perturbations in cavity solution}
Following the same procedure as in \cite{advani2018statistical}, we introduce the following susceptibilities:
\begin{eqnarray}
\chi^R_{\alpha \beta}&=&-\frac{\partial \bar{R}_\alpha}{\partial \omega_\beta}\\
\chi^N_{i \alpha}&=&-\frac{\partial \bar{N}_i}{\partial \omega_\alpha}\\ 
\nu^R_{\alpha i}&=&\frac{\partial \bar{R}_\alpha }{\partial m_i}\\
\nu^N_{i j}&=&\frac{\partial \bar{N}_i }{\partial m_j}
\end{eqnarray}
where we denote $\bar{X}$ as the steady-state value of $X$. Recall that the goal is to derive a set of self-consistency equations that relates the ecological system characterized by $M+1$ resources (variables) and $S+1$ species (constraints) to that with the new species and new resources removed: $(S+1, M+1)\rightarrow (S,M)$. To simplify notation, let $\bar{X}_{\backslash 0}$ denote the steady-state value of quantity $X$ in the absence of the new resource and new species. Since the introduction of a new species and resource represents only a small (order $1/M$) perturbation to the original ecological system, we can express the steady-state species and resource abundances in the $(S+1, M+1)$ system with a first-order Taylor expansion around the $(S,M)$ values. We note that the new terms $\sigma_c d_{i0}R_0$ in Eq. eq. (\ref{eq-n}) and $\sigma_c d_{0\alpha}N_0$ in eq. (\ref{eq-r}) can be treated as perturbations to $m_i$, and $K_\alpha$, respectively, yielding:
\begin{eqnarray}\label{eq:lcav}
\bar{N}_i = \bar{N}_{i /0} -\sigma_c\sum_{\beta /0} \chi^N_{i\beta} d_{0\beta}\bar{N}_0 - \sigma_c\sum_{j /0}\nu^N_{ij} d_{j0}  \bar{R}_0
\end{eqnarray}

\begin{eqnarray}\label{eq:Rcav}
\bar{R}_\alpha = \bar{R}_{\alpha /0} -\sigma_c\sum_{\beta /0} \chi^R_{\alpha  \beta} d_{0\beta}\bar{N}_0 - \sigma_c\sum_{j /0}\nu^R_{\alpha j} d_{j0}  \bar{R}_0
\end{eqnarray}
Note  $\sum_{j /0}$ and $\sum_{\beta /0}$ mean the sum excludes the new species 0 and the new resource 0. The next step is to plug eq. (\ref{eq:lcav}) and eq. (\ref{eq:Rcav}) into eq. (\ref{dy-N0}) and eq. (\ref{dy-R0}) and solve for the steady-state value of $N_0$ and $R_0$. 

\subsection{Self-consistency equations for species}
For the new cavity species, the steady equation takes the form
\begin{eqnarray}
0=\bar{N}_0 \left[ \mu\left< R\right>-m-\sigma_c^2\bar{N}_0\sum_{\alpha/0, \beta/0}\chi^R_{\alpha\beta}d_{0\alpha}d_{0\beta}-\sigma_c^2\bar{R}_0\sum_{\beta /0, j /0}\nu^R_{\beta j}d_{0\beta}d_{0 j}+\sum_{\beta /0}\sigma_c d_{0\beta} \bar{R}_{\beta/0}+\sigma_c d_{00} \bar{R}_{0}-\delta m_0 \right]
\end{eqnarray}
Notice that each of the sums in this equation is the sum over a large number of weak correlated random variables, and can therefore be well approximated by Gaussian random variables for large enough $M$ and $S$.   We can calculate the sum of the random variables:
\begin{eqnarray}
\sum_{\beta /0, j /0}\nu^R_{\beta j}d_{0\beta}d_{0 j} =  \frac{1}{M}\sum_{\beta /0, j /0} \nu^R_{\beta j} \delta_{ j 0}\delta_{ \beta 0}=0
\end{eqnarray}
\begin{eqnarray}
\sum_{\alpha/0 ,\beta /0}\chi^R_{\alpha\beta}d_{0\alpha}d_{0\beta} =  \frac{1}{M}\sum_{\alpha /0 ,\beta /0}\chi^R_{\alpha\beta} \delta_{\alpha\beta}=\frac{1}{M}\sum_\alpha \chi^R_{\alpha\alpha}=\frac{1}{M}\text{Tr}(\chi^R_{\alpha\beta}) =\chi
\end{eqnarray}
where $\chi$ is the average susceptibility.  Using these observations about above sums, we obtain
\begin{eqnarray}
0= \bar{N}_0 \left[ \mu\left< R\right>-m-\sigma_c^2\chi \bar{N}_0+\sum_{\beta /0}\sigma_c d_{0\beta} \bar{R}_{\beta/0}+\sigma_c d_{00} \bar{R}_{0}-\delta m_0 \right]+ \mathcal{O}(M^{-1/2}),\label{sol-N0}
\end{eqnarray}
Employing the Central Limit Theorem, we introduce an auxiliary Gaussian variable $z_N$ with zero mean and unit variance and rewrite this as
\begin{eqnarray}\label{z_N}
\sum_{\beta /0}\sigma_c d_{0\beta} \bar{R}_{\beta/0}+\sigma_c d_{0\beta} \bar{R}_{0}-\delta m_0 = z_N\sqrt{\sigma_c^2 q_R+\sigma_m^2} ,
 \end{eqnarray}
where $q_R$ is the second moment of the resource distribution, 
$$q_R=\frac{1}{M}\sum_\beta R_\beta^2.$$
We can solve eq. (\ref{sol-N0}) in terms of the quantities just defined:
\begin{eqnarray}
 \mu\left< R\right>-m-\sigma_c^2\chi \bar{N}_0+\sqrt{\sigma_c^2 q_R+\sigma_m^2}z_N \le 0  
 \end{eqnarray}
Inverting this equation one gets the steady state of species
\begin{eqnarray}\label{N0}
\bar{N}_0=\mathrm{max}\left[ 0, \quad \frac{\mu\left< R\right>-m+\sqrt{\sigma_c^2 q_R+\sigma_m^2}z_N}{\sigma_c^2 \chi}\right]
 \end{eqnarray}
which is a truncated Gaussian. 

Let $y=\text{max}\left(0, \frac{a}{b}+\frac{c}{b}z\right)$, with $z$ being a Gaussian random variable with zero mean and unit variance. Then its $j$-th moment is given by
\begin{eqnarray}
\left<y^j \right>&=&\frac{1}{\sqrt{2\pi}}\int_{-\frac{a}{c}}^\infty e^{-\frac{x^2}{2}}\left (\frac{c}{b}x+\frac{a}{b} \right)^j dx\\
&=&\left(\frac{c}{b}\right)^j\frac{1}{\sqrt{2\pi}}\int_{-\frac{a}{c}}^\infty e^{-\frac{x^2}{2}}\left (x+\frac{a}{c} \right)^j dx\\
&=&\left(\frac{c}{b}\right)^jw_j(\frac{a}{c})
\end{eqnarray}
here we define $w_j(\frac{a}{c})=\frac{1}{\sqrt{2\pi}}\int_{-\frac{a}{c}}^\infty e^{-\frac{x^2}{2}}\left (x+\frac{a}{c} \right)^j dx$

With this we can easily write down the self-consistency equations for the fraction of non-zero species and resources as well as the moments of their abundances at the steady state:
\begin{eqnarray}
\phi_N &=&\frac{S^*}{S}=  w_0\left(\frac{\mu\left< R\right>-m}{\sqrt{\sigma_c^2 q_R+\sigma_m^2}}\right)\label{phiN1}\\
 \left< N\right>&=&\frac{1}{S}\sum_{j}N_j=\left( \frac{\sqrt{\sigma_c^2 q_R+\sigma_m^2}}{\sigma_c^2 \chi}\right) w_1(\frac{ \mu\left< R\right>-m}{\sqrt{\sigma_c^2 q_R+\sigma_m^2}})\\
 q_N&=&\frac{1}{S}\sum_{j}N^2_j=\left( \frac{\sqrt{\sigma_c^2 q_R+\sigma_m^2}}{\sigma_c^2 \chi}\right)^2 w_2(\frac{ \mu\left< R\right>-m}{\sqrt{\sigma_c^2 q_R+\sigma_m^2}})\label{qN2}
\end{eqnarray}
Note that $S^*$ is the number of surviving species at the steady state.
\subsection{Self-consistency equations for resources}
We now derive the equations for the steady-state of the resource dynamics. Inserting eq. (\ref{eq:Rcav}) into eq. (\ref{dy-R0})  gives:
\begin{eqnarray}
0\!=\!K\!+\!\delta K_0\!-
\!\bar{R}_0 \left[ \omega+\gamma^{-1}\mu\left<N \right>\!-\!\sigma_c^2\bar{N}_0\!\!\!\!\sum_{\beta/0, j/0}\!\!\!\!\chi^N_{j\beta}d_{j 0}d_{0\beta}\!-\!\sigma_c^2\bar{R}_0\!\!\!\!\!\sum_{i /0, j /0}\!\!\!\!\nu^N_{i j}d_{0i}d_{0 j}\!+\!\sum_{j /0}\sigma_c d_{j0} \bar{N}_{j/0}\!+\!\sigma_c d_{00} \bar{N}_{0}\!+\!\delta \omega_0 \right]
\end{eqnarray}
We can simplify the sums by averaging over the random variables:
\begin{eqnarray}
\sum_{\beta/0, j/0} \chi^N_{j\beta}d_{j 0}d_{0\beta}=  \frac{1}{M}\sum_{\beta /0, j /0} \chi^N_{j\beta} \delta_{ j 0}\delta_{ \beta 0}=0
\end{eqnarray}
\begin{eqnarray}
\sum_{i /0, j /0} \nu^N_{i j}d_{0i}d_{0 j}=  \frac{1}{M}\sum_{i /0, j /0} \nu^N_{i j} \delta_{ij}=\frac{1}{M}\sum_i \nu^N_{i i}=\frac{1}{M}\text{Tr}(\nu^N_{ij}) =\gamma^{-1}\nu
\end{eqnarray}
where $\nu$ is the average susceptibility.  Finally, note that we can write
\begin{equation}\label{z_R}
\delta \omega_0 +\sum_j \sigma_c d_{j0}N_j = z_R \sqrt{\gamma^{-1}\sigma_c^2q_N+\sigma_\omega^2},
\end{equation}
where we have introduced another auxiliary Gaussian variable $z_R$ with zero mean and unit variance and $q_N$ is the second moment of the resource distribution defined in eq. (\ref{qN1}), 
Using these observations, we obtain a quadratic expression for the resource.
\begin{eqnarray} \label{1R0}
K+\delta K_0-(\omega_0+\gamma^{-1}\mu\left<N \right>+ \sqrt{\gamma^{-1}\sigma_c^2q_N+\sigma_\omega^2}   z_R)\bar{R}_0+\gamma^{-1}\sigma_c^2\nu \bar{R}_0^2=0 
\end{eqnarray}

\subsubsection{Cavity solution: without backreaction}
As discussed in the main text, we cannot solve the full resource equations exactly. For this reason, we perform an expansion, as a start, we calculate this equation by setting $\nu=0$ in the resource equation. This is equivalent in the TAP language of ignoring the backreaction term.

Under this assumption,
the quadratic equation for the resource, simply becomes a linear equation that can be re-arranged to give
\begin{equation}
\bar{R}_\alpha=\frac{K + \delta K_\alpha }{\omega+\gamma^{-1}\mu\left< N\right>+  z_R \sqrt{\gamma^{-1}\sigma_c^2q_N+\sigma_\omega^2}}\label{ralpha}
\end{equation}
Assuming the fluctuations in the denominator is small, $i.e.$ $\sqrt{\gamma^{-1}\sigma_c^2q_N+\sigma_\omega^2} \ll \omega+\gamma^{-1}\mu\left<N \right>$,  we can do a first-order Taylor expansion around the mean value and also ignore the coupling term between $\delta K_\alpha$ and $z_R$:
\begin{equation}
\bar{R}_\alpha=\frac{K + \delta K_\alpha }{\omega+\gamma^{-1}\mu\left< N\right>}-\frac{K \sqrt{\gamma^{-1}\sigma_c^2q_N+\sigma_\omega^2}  }{(\omega+\gamma^{-1}\mu\left< N\right>)^2}z_R
\end{equation}
With all these approximations, we get the first two moments of the steady-state resource abundance distribution:
\begin{equation}
\left<R\right>=\frac{K}{\omega+\gamma^{-1}\mu\left<N \right>}
\end{equation}
\begin{equation}
q_R=\left<R\right>^2+\frac{\sigma_K^2}{(\omega+\gamma^{-1}\mu\left<N \right>)^2}+\frac{K^2 (\gamma^{-1}\sigma_c^2q_N+\sigma_\omega^2)  }{(\omega+\gamma^{-1}\mu\left< N\right>)^4}
\end{equation}

The susceptibility is given by:
\begin{eqnarray}
\chi&=&-\left<\frac{\partial\bar{R}_\alpha }{\partial w_\alpha}\right>
=\left<\frac{K_\alpha}{(\omega_\alpha+\sum_j c_{j\alpha}\bar{N}_j)^2}+\frac{2K \sqrt{\gamma^{-1}\sigma_c^2q_N+\sigma_\omega^2}  }{(\omega+\gamma^{-1}\mu\left< N\right>)^3}z_R\right>
=\frac{K}{(\omega+\gamma^{-1}\mu\left<N \right>)^2}
\end{eqnarray}
Combined with self-consistency equations for species, we get the full set of :
\begin{eqnarray}
\phi_N =  w_0\left(\frac{\mu\left< R\right>-m}{\sqrt{\sigma_c^2 q_R+\sigma_m^2}}\right), &\quad& \chi=\frac{K}{(\omega+\gamma^{-1}\mu\left<N \right>)^2} \\
 \left< N\right>=\left( \frac{\sqrt{\sigma_c^2 q_R+\sigma_m^2}}{\sigma_c^2 \chi}\right) w_1(\frac{ \mu\left< R\right>-m}{\sqrt{\sigma_c^2 q_R+\sigma_m^2}}), 
 &\quad& 
 \left<R\right>=\frac{K}{\omega+\gamma^{-1}\mu\left<N \right>}\\
 q_N=\left( \frac{\sqrt{\sigma_c^2 q_R+\sigma_m^2}}{\sigma_c^2 \chi}\right)^2 w_2(\frac{ \mu\left< R\right>-m}{\sqrt{\sigma_c^2 q_R+\sigma_m^2}}),
 &\quad& 
 q_R=\left<R\right>^2+\frac{\sigma_K^2}{(\omega+\gamma^{-1}\mu\left<N \right>)^2}+\frac{K^2 (\gamma^{-1}\sigma_c^2q_N+\sigma_\omega^2)  }{(\omega+\gamma^{-1}\mu\left< N\right>)^4}
\end{eqnarray}
\subsubsection{Cavity solution: with backreaction correction}
We start again with the full resource equation:
\begin{eqnarray} \label{1R0}
K+\delta K_0-(\omega_0+\gamma^{-1}\mu\left<N \right>+ \sqrt{\gamma^{-1}\sigma_c^2q_N+\sigma_\omega^2}   z_R)\bar{R}_0+\gamma^{-1}\sigma_c^2\nu \bar{R}_0^2=0 
\end{eqnarray}
Since $R_0>0$ and $\nu<0$, the solution of eq. (\ref{1R0}) gives:

\begin{equation}\label{RFULL}
R_0=\frac{\omega+\gamma^{-1}\mu  \left<N\right>+\sqrt{\gamma^{-1}\sigma_c^2q_N+\sigma_\omega^2}   z_R }{2\gamma^{-1}\sigma_c^2\nu }- \frac{\sqrt{(\omega+\gamma^{-1}\mu\left<N\right>+\sqrt{\gamma^{-1}\sigma_c^2q_N+\sigma_\omega^2}z_R)^2-4\gamma^{-1}\nu \sigma_c^2(K+\delta K_0)}}{2\gamma^{-1} \sigma_c^2\nu }
\end{equation}

For the  $1^{st}$ order expansion, we assume  $4\gamma^{-1}\nu \sigma_c^2 \delta K_0 +2\sqrt{\gamma^{-1}\sigma_c^2q_N+\sigma_\omega^2}z_R+(\gamma^{-1}\sigma_c^2q_N+\sigma_\omega^2)z_R^2\ll( \omega+\gamma^{-1}\mu\left<N \right>)^2-4\gamma^{-1}\nu \sigma_c^2K$ and do a 1st order expansion around the mean of the form:
\begin{eqnarray}
&&\sqrt{(\omega+\gamma^{-1}\mu\left<N\right>+\sqrt{\gamma^{-1}\sigma_c^2q_N+\sigma_\omega^2}z_R)^2-4\gamma^{-1}\nu \sigma_c^2(K+\delta K_0)}\nonumber\\
=&&\sqrt{(\omega+\gamma^{-1}\mu\left<N\right>)^2-4\gamma^{-1}\nu \sigma_c^2K}+\frac{(\gamma^{-1}\sigma_c^2q_N+\sigma_\omega^2)z_R^2+ 2(\omega+\gamma^{-1}\mu\left<N\right>)\sqrt{\gamma^{-1}\sigma_c^2q_N+\sigma_\omega^2}z_R-4\gamma^{-1}\nu \sigma_c^2\delta K_0}{2\sqrt{(\omega+\gamma^{-1}\mu\left<N\right>)^2-4\gamma^{-1}\nu \sigma_c^2K}}
\end{eqnarray}

Using these expressions, the moments of their abundances at  steady state can be calculated yielding:

\begin{equation}\label{R01}
\left< R \right>=\frac{\omega+\gamma^{-1}\mu\left<N\right> }{2\gamma^{-1}\sigma_c^2\nu }- \frac{\sqrt{(\omega+\gamma^{-1}\mu\left<N\right>)^2 -4\gamma^{-1}\nu \sigma_c^2K }}{2\gamma^{-1} \sigma_c^2\nu }-\frac{\gamma^{-1}\sigma_c^2q_N+\sigma_\omega^2}{4\gamma^{-1} \sigma_c^2\nu \sqrt{(\omega+\gamma^{-1}\mu\left<N\right>)^2-4\gamma^{-1}\nu \sigma_c^2K}}
\end{equation}
\begin{eqnarray}
q_R&=&\left< R \right>^2
+\frac{(\gamma^{-1}\sigma_c^2q_N+\sigma_\omega^2 )^2 + 8(\gamma^{-1}\nu \sigma_c^2\sigma_K)^2}{2(2\gamma^{-1}\sigma_c^2\nu)^2[(\omega+\gamma^{-1}\mu\left<N\right>)^2-4\gamma^{-1}\nu \sigma_c^2K]}\nonumber\\
&+&\frac{(\gamma^{-1}\sigma_c^2q_N+\sigma_\omega^2)[\sqrt{(\omega+\gamma^{-1}\mu\left<N\right>)^2-4\gamma^{-1}\nu \sigma_c^2K}-(\omega+\gamma^{-1}\mu\left<N\right>)]^2 }{(2\gamma^{-1}\sigma_c^2\nu)^2[(\omega+\gamma^{-1}\mu\left<N\right>)^2-4\gamma^{-1}\nu \sigma_c^2K]} 
\end{eqnarray}
From eq. (\ref{RFULL}),\label{ChiFULL}
\begin{eqnarray}\label{chibe}
\frac{\partial R_0 }{\partial \omega}
=\frac{1 }{2\gamma^{-1}\sigma_c^2\nu }\left\{1- \frac{\omega+\gamma^{-1}\mu\left<N\right>+\sqrt{\gamma^{-1}\sigma_c^2q_N+\sigma_\omega^2}z_R}{\sqrt{(\omega+\gamma^{-1}\mu\left<N\right>+\sqrt{\gamma^{-1}\sigma_c^2q_N+\sigma_\omega^2}z_R)^2-4\gamma^{-1}\nu \sigma_c^2(K+\delta K_0)} }\right\}
\end{eqnarray}
The term inside the bracket can be expanded as:
\begin{eqnarray}
&&\frac{\omega+\gamma^{-1}\mu\left<N\right>+\sqrt{\gamma^{-1}\sigma_c^2q_N+\sigma_\omega^2}z_R}{\sqrt{(\omega+\gamma^{-1}\mu\left<N\right>+\sqrt{\gamma^{-1}\sigma_c^2q_N+\sigma_\omega^2}z_R)^2-4\gamma^{-1}\nu \sigma_c^2(K+\delta K_0)} }\\
&\approx&\resizebox{.9\hsize}{!}
{$ \frac{\omega+\gamma^{-1}\mu\left<N\right>+\sqrt{\gamma^{-1}\sigma_c^2q_N+\sigma_\omega^2}z_R}{\sqrt{(\omega+\gamma^{-1}\mu\left<N\right>)^2-4\gamma^{-1}\nu \sigma_c^2 K} }\left[1-\frac{(\gamma^{-1}\sigma_c^2q_N+\sigma_\omega^2)z_R^2+ 2(\omega+\gamma^{-1}\mu\left<N\right>)\sqrt{\gamma^{-1}\sigma_c^2q_N+\sigma_\omega^2}z_R-4\gamma^{-1}\nu \sigma_c^2\delta K_0}{2(\omega+\gamma^{-1}\mu\left<N\right>)^2-4\gamma^{-1}\nu \sigma_c^2K}\right]$}\nonumber
\end{eqnarray}

The susceptibilities are given by averaging eq. (\ref{chibe})
\begin{eqnarray}
\chi&=& -\left<\frac{\partial R}{\partial \omega} \right>\\
&=&-\frac{1}{2\gamma^{-1}\nu \sigma_c^2}\left\{1-\frac{\omega+\gamma^{-1}\mu\left<N \right>}{\sqrt{(\omega+\gamma^{-1}\mu\left<N \right>)^2-4\gamma^{-1}\nu \sigma_c^2 K}}
+\frac{3(\gamma^{-1}\sigma_c^2q_N+\sigma_\omega^2)(\omega+\gamma^{-1}\mu\left<N \right>)}{2[(\omega+\gamma^{-1}\mu\left<N\right>)^2-4\gamma^{-1}\nu \sigma_c^2K]^{3/2}}\right\}\\
\nu&=&\left<\frac{\partial N}{\partial m} \right>= -\frac{\phi_N}{\sigma_c^2 \chi} 
\end{eqnarray}


Combined with self-consistency equations for species, get the full set of $1^{st}$ order self-consistency equations:
\begin{eqnarray}
\phi_N &=& w_0\left(\frac{\mu\left< R\right>-m}{\sqrt{\sigma_c^2 q_R+\sigma_m^2}}\right)\\
 \left< N\right>&=&\left( \frac{\sqrt{\sigma_c^2 q_R+\sigma_m^2}}{\sigma_c^2 \chi}\right) w_1(\frac{ \mu\left< R\right>-m}{\sqrt{\sigma_c^2 q_R+\sigma_m^2}})\label{N1}\\
 q_N&=&\left( \frac{\sqrt{\sigma_c^2 q_R+\sigma_m^2}}{\sigma_c^2 \chi}\right)^2 w_2(\frac{ \mu\left< R\right>-m}{\sqrt{\sigma_c^2 q_R+\sigma_m^2}})\label{qN1}
\end{eqnarray}

\begin{equation}
\left< R \right>=\frac{\omega+\gamma^{-1}\mu\left<N\right> }{2\gamma^{-1}\sigma_c^2\nu }- \frac{\sqrt{(\omega+\gamma^{-1}\mu\left<N\right>)^2 -4\gamma^{-1}\nu \sigma_c^2K }}{2\gamma^{-1} \sigma_c^2\nu }-\frac{\gamma^{-1}\sigma_c^2q_N+\sigma_\omega^2}{4\gamma^{-1} \sigma_c^2\nu \sqrt{(\omega+\gamma^{-1}\mu\left<N\right>)^2-4\gamma^{-1}\nu \sigma_c^2K}}
\end{equation}
\begin{eqnarray}
q_R&=&\left< R \right>^2
+\frac{(\gamma^{-1}\sigma_c^2q_N+\sigma_\omega^2 )^2 + 8(\gamma^{-1}\nu \sigma_c^2\sigma_K)^2}{2(2\gamma^{-1}\sigma_c^2\nu)^2[(\omega+\gamma^{-1}\mu\left<N\right>)^2-4\gamma^{-1}\nu \sigma_c^2K]}\nonumber\\
&+&\frac{(\gamma^{-1}\sigma_c^2q_N+\sigma_\omega^2)[\sqrt{(\omega+\gamma^{-1}\mu\left<N\right>)^2-4\gamma^{-1}\nu \sigma_c^2K}-(\omega+\gamma^{-1}\mu\left<N\right>)]^2 }{(2\gamma^{-1}\sigma_c^2\nu)^2[(\omega+\gamma^{-1}\mu\left<N\right>)^2-4\gamma^{-1}\nu \sigma_c^2K]} 
\end{eqnarray}
\begin{eqnarray}
\chi&=& -\frac{1}{2\gamma^{-1}\nu \sigma_c^2}\left\{1-\frac{\omega+\gamma^{-1}\mu\left<N \right>}{\sqrt{(\omega+\gamma^{-1}\mu\left<N \right>)^2-4\gamma^{-1}\nu \sigma_c^2 K}}
+\frac{3(\gamma^{-1}\sigma_c^2q_N+\sigma_\omega^2)(\omega+\gamma^{-1}\mu\left<N \right>)}{2[(\omega+\gamma^{-1}\mu\left<N\right>)^2-4\gamma^{-1}\nu \sigma_c^2 K]^{3/2}}\right\}\\\label{chifllexp}
\nu&=& -\frac{\phi_N}{\sigma_c^2 \chi} \label{chinu}
\end{eqnarray}

\section{Comparison between with and without backreaction}\label{compare01}
We can reduce the cavity solution with backreaction to the simpler one when $\sigma_c$ is large. In fact all the complexity of cavity solution with backreaction comes from the expression for eq. (\ref{RFULL}):
\begin{equation}
R_0=\frac{\omega+\gamma^{-1}\mu\left<N\right>+\sqrt{\gamma^{-1}\sigma_c^2q_N+\sigma_\omega^2}   z_R }{2\gamma^{-1}\sigma_c^2\nu }- \frac{\sqrt{(\omega+\gamma^{-1}\mu\left<N\right>+\sqrt{\gamma^{-1}\sigma_c^2q_N+\sigma_\omega^2}z_R)^2-4\gamma^{-1}\nu \sigma_c^2(K+\delta K_0)}}{2\gamma^{-1} \sigma_c^2\nu }
\end{equation}
However, if we assume $(\omega+\gamma^{-1}\mu\left<N\right>+\sqrt{\gamma^{-1}\sigma_c^2q_N+\sigma_\omega^2}z_R)^2 \gg -4\gamma^{-1}\nu \sigma_c^2(K+\delta K_0)$, we can expand the second term following $\sqrt{1-x}\approx 1-\frac{x}{2}-\frac{x^2}{8}+\mathcal{O}(x^3)$.
\begin{equation}
R_0=\frac{K+\delta K_0}{\omega+\gamma^{-1}\mu\left<N\right>+\sqrt{\gamma^{-1}\sigma_c^2q_N+\sigma_\omega^2}z_R}+\frac{\gamma^{-1}\sigma_c^2\nu (K+\delta K_0)^2}{(\omega+\gamma^{-1}\mu\left<N\right>+\sqrt{\gamma^{-1}\sigma_c^2q_N+\sigma_\omega^2}z_R)^3}\label{R01st}
\end{equation}
The first term of above equation is the cavity solution without backreaction.

\subsection{ Comparing the cavity solutions to numerical simulations}
\begin{figure}[h]
\centering
\includegraphics[width=1.\textwidth]{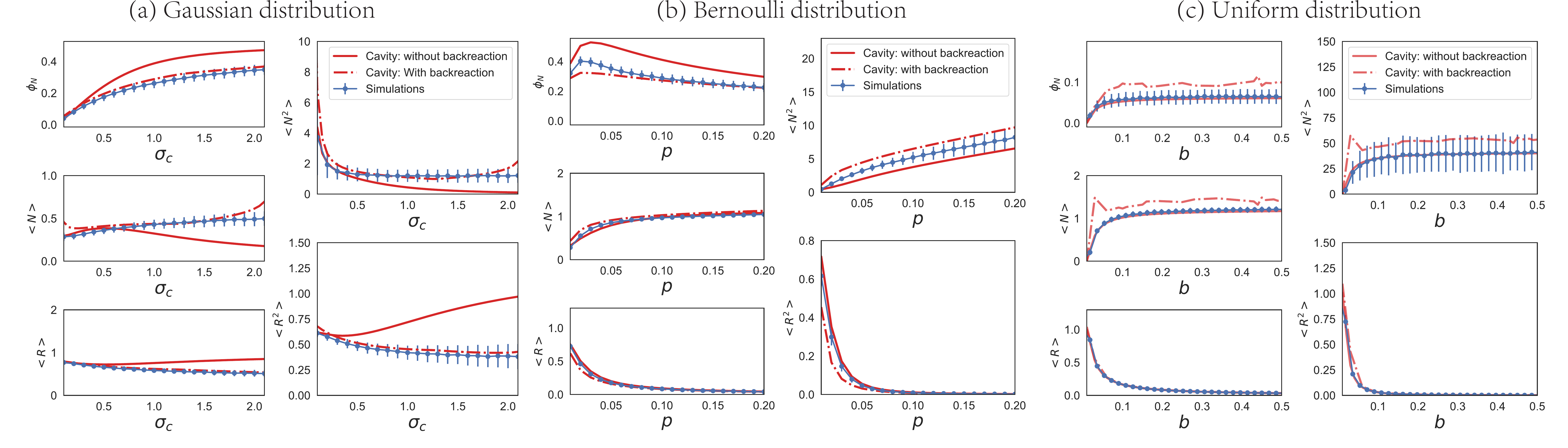}
\caption{Comparison of numerics and cavity solutions with and without the backreaction term as a function of $\sigma_c$. $\phi_N={S^* \over S}$ is the fraction of surviving species. $\left< N \right>, \left< N^2 \right>,\left< R \right>$ and $\left< R^2 \right>$ are the first and second moments of the species and resources distribution respectively. The simulations details can be found at the SM: \ref{simulation}.  $\mathbf{C}$  is sampled either from a Gaussian, Bernoulli, or uniform distribution as indicated.}
\label{other}
\end{figure}

We show a comparison between theoretical and numerical results for different choices of how to sample the consumption matrix in Fig. 2 in the main text and Fig. \ref{other}. These figures show that the cavity solution with backreaction performs better for the Gaussian and Bernoulli cases. However, in the uniform case, the cavity solution without backreaction matches with numerical simulations perfectly, while the cavity solution with backreaction performs worse than without backreaction.  In the section \ref{compare01}, we have shown the cavity solution with backreaction  can be reduced to the cavity solution without backreaction and hence should be a more robust solution. So why does it perform badly in the uniform case? The reason is that in the uniform case $\mu=Mb/2 \gg 1$ when the system size $M$ is large, leading to $|\chi | \sim \frac{1}{(\omega+
\gamma^{-1}\mu\left< N\right>)^2} \ll 1$. From eqs. (\ref{N1}, \ref{qN1}), we see that both $\left< N\right>$ and $\left< N^2\right>$ depends on $\frac{1}{\chi} \gg 1$ and the numerical solver becomes unstable.

\section{Simulation Details}\label{simulation}
\subsection{Parameters}
All simulations are done with the CVXPY package\cite{cvxpy_rewriting}
in PYTHON 3. All codes are available on GitHub at \url{https://github.com/Emergent-Behaviors-in-Biology/species-packing-bound}.

\begin{itemize}
\item Fig. 2: the consumer matrix $\mathbf{C}$ is sampled from the Gaussian distribution $\mathcal{N}(\frac{\mu}{M},\frac{\sigma_c}{\sqrt{M}} )$. $S=100$, $M=100$, $\mu=1$, $K=1$, $\sigma_K=0.1$ ,  $m=1$, $\sigma_m=0.1$, $\omega=1$, $\sigma_\omega=0$  and each data point is averaged from 1000 independent realizations.  We only provide the cavity solution with backreaction here.

\item Fig. 3, Fig. \ref{Nsim}, Fig. \ref{rdistribution}, Fig. \ref{gamma}: the consumer matrix $\mathbf{C}$ is sampled from the Gaussian distribution $\mathcal{N}(\frac{\mu}{M},\frac{\sigma_c}{\sqrt{M}} )$. $S=500$, $M=100$, $\mu=1$, $\sigma_\kappa=0.1$ ,  $m=1$, $\sigma_m=0.1$, $\omega=1$, $\sigma_\omega=0$  for externally supplied resource dynamics and $S=500$, $M=100$, $\mu=1$, $\sigma_\kappa=0.1$ ,  $m=1$, $\sigma_m=0.1$, $\tau=1$, $\sigma_\tau=0$ for the self-renewing one. Each data point is averaged from 1000 independent realizations. For Fig. \ref{Nsim} , $K=10$.  For Fig. \ref{rdistribution}, $\sigma_c=5$, $K$ and $\kappa$ are fixed at 4; For Fig. \ref{gamma}, $\sigma_c=5$, $\kappa=4$, $S/M$ has a range from 1 to 100, and each data point is averaged from 100 independent realizations.

\item Fig. 4: the consumer matrix $\mathbf{C}$ is sampled from the Bernoulli distribution $\mathit{Bernoulli}(p)$ and $p$ are fixed to 0.1, 0.2 and 0.1. $m_i$ follows metabolic
tradeoffs Eq. (\ref{2ndconst}) with $\sigma_\epsilon=0$, $\tilde{m}=1$. We also set $S=500$, $M=100$, $K=10$, $\sigma_K=0.1$. Each data point is averaged from 100 independent realizations. 

\item Fig. \ref{other}(a): the simulation is the same as Fig. 2. We show both the cavity solutions with and without reaction here.

\item Fig. \ref{other}(b): the consumer matrix $\mathbf{C}$ is sampled from the Bernoulli distribution $\mathit{Bernoulli}(p)$. $S=100$, $M=100$, $K=1$, $\sigma_K=0.1$ ,  $m=1$, $\sigma_m=0.1$, $\omega=1$, $\sigma_\omega=0$  and each data point is averaged from 1000 independent realizations. The cavity solution is obtained by approximating the Bernoulli distribution to the corresponding Gaussian distribution $i.e.$ $\mu=pM$, $\sigma_c=\sqrt{Mp(1-p)}$

\item Fig. \ref{other}(c): the consumer matrix $\mathbf{C}$ is sampled from the uniform distribution $\mathcal{U}(0, b)$. $S=100$, $M=100$, $K=1$, $\sigma_K=0.1$ ,  $m=1$, $\sigma_m=0.1$, $\omega=1$, $\sigma_\omega=0$ and each data point is averaged from 1000 independent realizations. The cavity solution is obtained by approximating the uniform distribution to the corresponding Gaussian distribution, $i.e.$ $\mu=bM/2$, $\sigma_c=b\sqrt{M/12}$.

\item Fig. \ref{otherbound}(a): the consumer matrix $\mathbf{C}$ is sampled from the Bernoulli distribution $\mathit{Bernoulli}(p)$. $S=500$, $M=100$, $K=1$, $\sigma_K=0.1$ ,  $m=1$, $\sigma_m=0.1$, $\omega=1$, $\sigma_\omega=0$  and each data point is averaged from 1000 independent realizations. The cavity solution is obtained by approximating the Bernoulli distribution to the corresponding Gaussian distribution $i.e.$ $\mu=pM$, $\sigma_c=\sqrt{Mp(1-p)}$

\item Fig. \ref{otherbound}(b): the consumer matrix $\mathbf{C}$ is sampled from the uniform distribution $\mathcal{U}(0, b)$. $S=500$, $M=100$, $K=1$, $\sigma_K=0.1$ ,  $m=1$, $\sigma_m=0.1$, $\omega=1$, $\sigma_\omega=0$ and each data point is averaged from 1000 independent realizations. The cavity solution is obtained by approximating the uniform distribution to the corresponding Gaussian distribution, $i.e.$ $\mu=bM/2$, $\sigma_c=b\sqrt{M/12}$.

\end{itemize}

\subsection{Distinction between extinct and surviving species}
\begin{figure}[h]
\centering
\includegraphics[width=1.\textwidth]{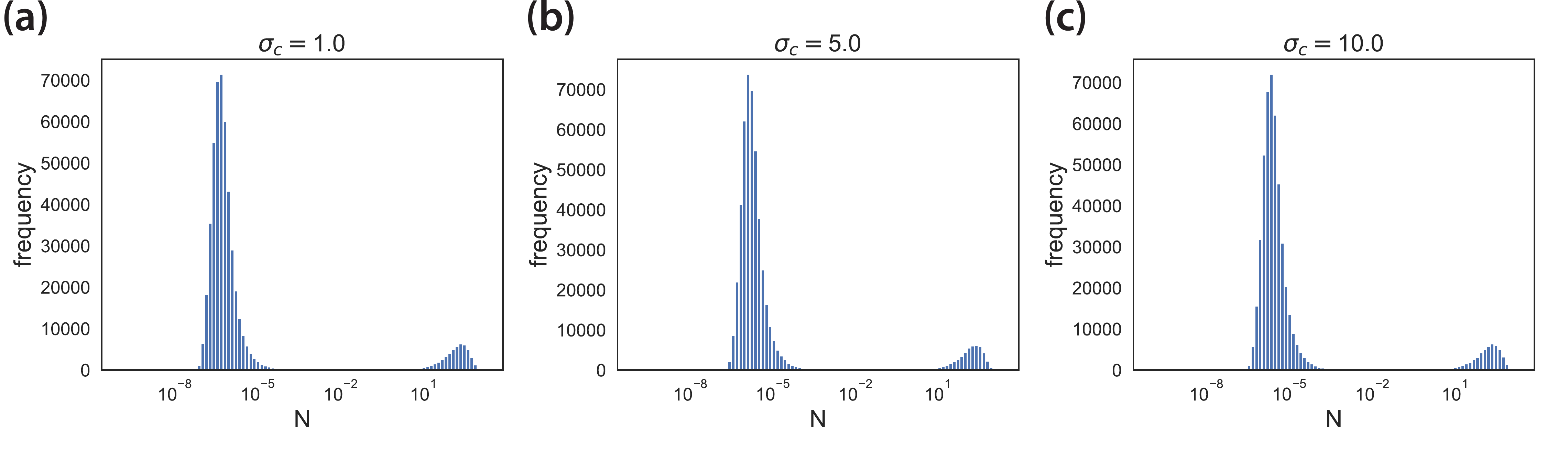}
\caption{Species abundance $N$ in equilibrium at different $\sigma_c$ for externally supplied resource dynamics at $K=10$. The simulations parameters can be found at the SM: \ref{simulation}.}
\label{Nsim}
\end{figure}

In the main text, we show that the value of species packing $\frac{S^*}{M}$ for the externally supplied resources must be smaller than 0.5. However, in numerical simulations,  even for the extinct  species  the abundance is never exactly equal 0 due to numerical errors.  As a result, we must choose a threshold to distinguish extinct and surviving species in order to calculate $S^*$. Since we are using the equivalence with convex optimization  to solve the generalized consumer-resource models\cite{mehta2019constrained, marsland2019minimum}, we can easily choose a reasonable threshold (e.g. $10^{-2}$ in Fig. \ref{Nsim}) since the extinct and surviving species are well separated in two peaks (see Fig. \ref{Nsim}).
\section{An upper bound for species packing}\label{bound}
By analyzing the susceptibilities in  the full Cavity solutions, an upper bound for species packing can be derived for both resource dynamics in GCRMs. The derivations can also be extended to the case where metabolic tradeoffs impose hard or soft constraints on the parameter values.
\subsection{Externally supplied resource dynamics }
The response functions $\chi$ and $\nu$ can be written as:
\begin{eqnarray}
\chi&=& -\frac{1 }{2\gamma^{-1}\sigma_c^2\nu }\left\{1-\left< \frac{\omega+\gamma^{-1}\mu\left<N\right>+\sqrt{\gamma^{-1}\sigma_c^2q_N+\sigma_\omega^2}z_R}{\sqrt{(\omega+\gamma^{-1}\mu\left<N\right>+\sqrt{\gamma^{-1}\sigma_c^2q_N+\sigma_\omega^2}z_R)^2-4\gamma^{-1}\nu \sigma_c^2(K+\delta K_0)} }\right>\right\}\label{chifll}\\
\nu&=& -\frac{\phi_N}{\sigma_c^2 \chi} \label{nufl2}
\end{eqnarray}
Substituting eq. (\ref{nufl2}) into eq. (\ref{chifll}) and rearranging yields
\begin{eqnarray}
\gamma^{-1}\phi_N= \frac{1 }{2}\left\{1-\left< \frac{\omega+\gamma^{-1}\mu\left<N\right>+\sqrt{\gamma^{-1}\sigma_c^2q_N+\sigma_\omega^2}z_R}{\sqrt{(\omega+\gamma^{-1}\mu\left<N\right>+\sqrt{\gamma^{-1}\sigma_c^2q_N+\sigma_\omega^2}z_R)^2-4\gamma^{-1}\nu \sigma_c^2(K+\delta K_0)} }\right>\right\}.
\end{eqnarray}
The numerator of the term in angle brackets is the total depletion rate for a given resource when it is first added to the system. Depletion rates are always positive in this model, so the right-hand side is always less than 1/2. Noticing  $\gamma=\frac{M}{S}$, $\phi_N=S^*/S$, $\chi>0$, we immediately obtain an upper bound on $\frac{S^*}{M}$:
\begin{equation}
\frac{1}{2} >\frac{S^*}{M}.
\end{equation}

\subsection{Self-renewing(MacArthur's) resource dynamics }
Using the analytical expressions $\chi$, $\nu$ and self-consistent equations in ref. \cite{cui2019diverse}, we can derive the following expressions:
\begin{eqnarray}
\left< N\right>=\left(\frac{\sqrt{\sigma_c^2 q_R+\sigma_m^2}}{\sigma_c^2(\phi_R-\gamma^{-1} \phi_N)}\right)w_1\left(\frac{\mu\left< R\right>-m}{\sqrt{\sigma_c^2 q_R+\sigma_m^2}}\right),\quad
\left< R\right>=\left(\frac{\sqrt{\gamma^{-1}\sigma_c^2 q_N+\sigma_K^2}}{\phi_R(\phi_R-\gamma^{-1} \phi_N)^{-1}}\right)w_1\left(\frac{\kappa-\gamma^{-1}\mu\left< N\right>}{\sqrt{\gamma^{-1}\sigma_c^2 q_N+\sigma_K^2}}\right)
\end{eqnarray}
To derive bounds, we consider various limits of these expressions.  First, consider the case were we put many species $S \rightarrow \infty$ into the ecosystem with fixed number of resources $M$,  (i.e $\gamma=\frac{M}{S} \rightarrow 0$). In order to keep $\left< N\right>$ positive, we must have $\phi_R-\gamma^{-1} \phi_N>0$, giving an upper bound:
\begin{eqnarray}
1 \ge \frac{M^*}{M}> \frac{S^*}{M}
\end{eqnarray}

\subsection{Externally supplied resources with metabolic tradeoffs}
Here we consider two kinds of constraints on the parameters, encoding metabolic tradeoffs. In the first, the maintenance cost $m_i = m$ is the same for all species, and the sum of the consumption preferences is constrained to equal some fixed ``enzyme budget'' $E$ that is nearly the same for all species:
\begin{align}
\sum_{\alpha} C_{i\alpha} = E + \delta E_i
\end{align}
where $\delta E_i$ is a small random variable with mean zero and variance $\sigma_E^2$. A hard constraint can be generated by taking $\sigma_E = 0$.

The second kind of constraint does not make any assumptions about $C_{i\alpha}$, but assigns a cost $\tilde{m}$ to every unit of consumption capacity, so that
\begin{align}\label{2ndconst}
m_i = (1+\epsilon_i)\tilde{m}\sum_\alpha C_{i\alpha}+\delta m_i
\end{align}
where $\epsilon_i$ and $\delta m_i$ are small random variables with mean zero and variances $\sigma_\epsilon^2$ and $\sigma_m^2$, respectively. A hard constraint can be generated by taking $\sigma_\epsilon = \sigma_m = 0$.

In the simplest way of setting up the first constraint, the equilibrium equations actually reduce to the same form as the second. Specifically, one usually generates a consumer preference matrix satisfying the constraint by first generating an i.i.d. matrix $\tilde{C}_{i\alpha}$, and then setting $C_{i\alpha} = (E+\delta E_i)\tilde{C}_{i\alpha}/\sum_\beta \tilde{C}_{i\beta}$. The resulting dynamics can be written as:

\begin{eqnarray}
\frac{dN_i}{dt} &=& N_i\left[\sum_\alpha  (E+\delta E_i)\frac{\tilde{C}_{i\alpha}}{\sum_\beta \tilde{C}_{i\beta}} R_\alpha - m_i\right]\\
&=& \frac{N_i (E+\delta E_i)}{\sum_\beta \tilde{C}_{i\beta}}\left[ \sum_\alpha \tilde{C}_{i\alpha} R_\alpha - m \frac{\sum_\beta \tilde{C}_{i\beta}}{E+\delta E_i}\right].
\end{eqnarray}

Dropping the tilde's, we can write the equilibrium condition in the same form that results from the second kind of constraint:
\begin{eqnarray}
0 = N_i \{ \sum_\alpha C_{i\alpha} [R_\alpha - (1+\epsilon_i) \tilde{m}] - \delta m_i \}
\label{eq:constrained}
\end{eqnarray}
with 
\begin{eqnarray}
\tilde{m} &=& \frac{m}{E}\\
\epsilon_i &=& -\frac{\delta E_i}{E}\\
\delta m_i &=& 0.
\end{eqnarray}

Inspection of Equation \ref{eq:constrained} immediately reveals an important novelty: now when we add a new resource as part of the cavity protocol, the perturbation to the growth rate can either be positive or negative, depending on the sign of $[R_\alpha - (1+\epsilon_i)\tilde{m}]$. This turns out to be the crucial factor that prevents the proof of the $S^*/M < 1/2$ bound from going through, regardless of the size of $\sigma_\epsilon$ or $\sigma_m$.

Following the same steps as above, we arrive at the following set of equilibrium conditions for the new species $N_0$ and resource $R_0$:
\begin{eqnarray}
0&=& \bar{N}_0\left[ \mu\langle R\rangle - \mu\tilde{m} + \sigma_N z_N - \sigma_c^2 \chi \bar{N}_0\right]\label{eq:N0}\\
0&=& K + \delta K_0 - (\omega + \gamma^{-1} \mu \langle N \rangle + \sigma_R z_R + \gamma^{-1} \sigma_c^2 \nu \tilde{m})\bar{R}_0 + \gamma^{-1} \sigma_c^2 \nu \bar{R}_0^2\label{eq:R0}
\end{eqnarray}
where
\begin{eqnarray}
\sigma_N^2 &=& \sigma_m^2 + \sigma_c^2 [q_R - 2 \tilde{m}\langle R\rangle + \tilde{m}^2(1+\sigma_\epsilon^2)]\\
\sigma_R^2 &=& \sigma_\omega^2 + \gamma^{-1}\sigma_c^2 q_N + \gamma^{-2}\sigma_c^4 \nu^2 \tilde{m}^2\sigma_\epsilon^2.
\end{eqnarray}
These are nearly identical to the equations we had before. The two key changes are the presence of a term with a negative sign inside the coefficient $\sigma_N$ of the random variable $z_N$, and the $\gamma^{-1} \sigma_c^2\nu\tilde{m}$ term inside the parentheses in the equation for the resources. 

We can now proceed in the same way as before, solving for $\bar{N}_0$ and $\bar{R}_0$ and taking derivatives to compute the susceptibilities. We find:
\begin{eqnarray}
\chi &=& -\frac{1}{2\gamma^{-1}\sigma_c^2 \nu} \left\{1- \left\langle \frac{\omega + \gamma^{-1}\mu \langle N\rangle + \sigma_R z_R + \gamma^{-1} \sigma_c^2 \nu\tilde{m}}{\sqrt{(\omega + \gamma^{-1}\mu \langle N\rangle + \sigma_R z_R + \gamma^{-1} \sigma_c^2 \nu\tilde{m})^2 - 4\gamma^{-1}\sigma_c^2\nu(K + \delta K_0)}}\right\rangle\right\}\label{eq:chi}\\
\nu &=& -\frac{\phi_N}{\sigma_c^2 \chi}\label{eq:nu}
\end{eqnarray}
This is almost the same as the expression in Equation (\ref{chifll}) obtained in the absence of constraints, except for the extra term $\gamma^{-1} \sigma_c^2 \nu\tilde{m}$ in the numerator and denominator. This term is significant because $\nu$ is a negative number, and if its absolute value is large enough, it can make the whole term in angle brackets negative. Inserting the second equation into the first, we obtain a formula for $S^*/M$:
\begin{eqnarray}
\frac{S^*}{M} = \gamma^{-1}\phi_N &= \frac{1}{2} \left\{1- \left\langle \frac{\omega + \gamma^{-1}\mu \langle N\rangle + \sigma_R z_R + \gamma^{-1} \sigma_c^2 \nu\tilde{m}}{\sqrt{(\omega + \gamma^{-1}\mu \langle N\rangle + \sigma_R z_R + \gamma^{-1} \sigma_c^2 \nu\tilde{m})^2 - 4\gamma^{-1}\sigma_c^2\nu(K + \delta K_0)}}\right\rangle\right\}
\end{eqnarray}
The term in brackets can now be negative, but is always greater than -1. We thus obtain the bound:
\begin{eqnarray}
\frac{S^*}{M} < 1.
\end{eqnarray}
The term approaches -1 in the limit $\nu\to - \infty$, which is the same limit required to saturate the bound in the model with self-renewing resources. As in that case, the limit cannot actually be achieved, because $\nu\to -\infty$ implies $\chi\to 0$ (Equation (\ref{eq:nu})), and $\chi$ appears in the denominator of the final expression for $\bar{N}_0$ (Equation (\ref{N0})), while the numerator always remains finite. 

\begin{figure}[t]
\centering
\includegraphics[width=0.8\textwidth]{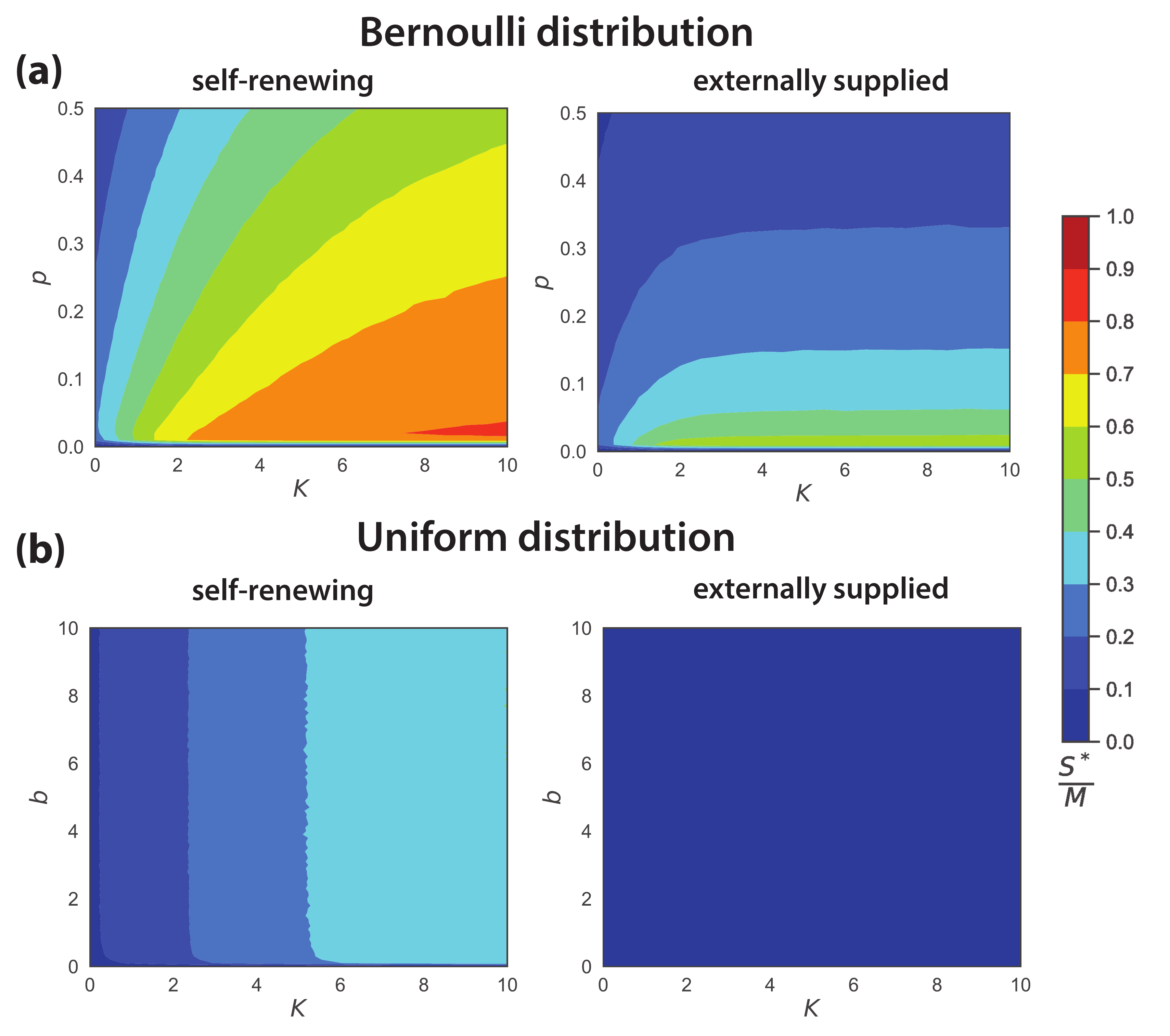}
\caption{Comparison of species packing $\frac{S^*}{M}$ for different distributions of consumption matrices $\mathbf{C}$ with self-renewing and externally-supplied resource dynamics. The simulations represent averages from 1000 independent realizations with the system size $M=100$, $S=500$ and parameters at the SM: \ref{simulation}.}
\label{otherbound}
\end{figure}

The only way to achieve the limit $\frac{S^*}{M} = 1$ is to make the numerator vanish in the same way as the denominator, which can only happen in the presence of hard constraints $\sigma_m = \sigma_\epsilon = 0$. In this case, it is easy to see that setting $R_\alpha = \tilde{m}$ for all $\alpha$ and $\chi\to 0, \nu \to -\infty$ solves both the steady state equations, regardless of the value of $\tilde{N}_0$. In Equation (\ref{eq:N0}) for $\tilde{N}_0$, the mean and the fluctuating part inside the brackets both vanish individually ($\mu\langle R\rangle - \mu\tilde{m}=0$, $\sigma_N = 0$), and the back-reaction term also vanishes ($\sigma_c^2 \chi \bar{N}_0 = 0$), leaving the equation trivially satisfied. In Equation (\ref{eq:R0}) only the terms with $\nu$ are significant in this limit, and they cancel each other perfectly. This is the ``shielded phase'' discussed in \cite{tikhonov2017collective}. 

Note also that if we take the $\chi\to 0$, $\nu \to -\infty$ limit first, before performing any substitutions, Equations (\ref{eq:chi}) and (\ref{eq:nu}) are satisfied independently of the choice of $\phi_N$. This means that $\gamma^{-1}\phi_N = S^*/M$ can be greater than 1, as observed in the simulations of \cite{posfai2017metabolic}. 

\subsection{Numerical evidence}
We show a comparison between the cavity solution and numerical results in Fig. 3 and Fig. \ref{otherbound} for three different distributions of the consumption matrix $\mathbf{C}$. For the Gaussian and Bernoulli distributions, $\frac{S^*}{M}$ can reach the upper bound we derived for two different resource dynamics. For externally supplied resource dynamics,  $\frac{S^*}{M}$ never exceeds 0.5.  For the uniform case, since the fluctuation of consumption matrix is small, the niche overlap is large and there is fierce competitions among species and theses ecosystems live very far from the upper bounds we derive. However, even for the uniform case, the species packing fraction is significantly larger for self-renewing resource dynamics than externally supplied resource dynamics. For the Bernoulli case, when the binomial probability $p\sim 1/M $, the bound can be slightly above 0.5, as shown in Fig. \ref{otherbound}.  In this regime, the consumer matrix is sparse. Each species only consumes one or two different resources and species rarely compete with each other making it is possible to pack more species. 

\subsection{Numerical analysis}
\begin{figure}
\centering
\includegraphics[width=0.95\textwidth]{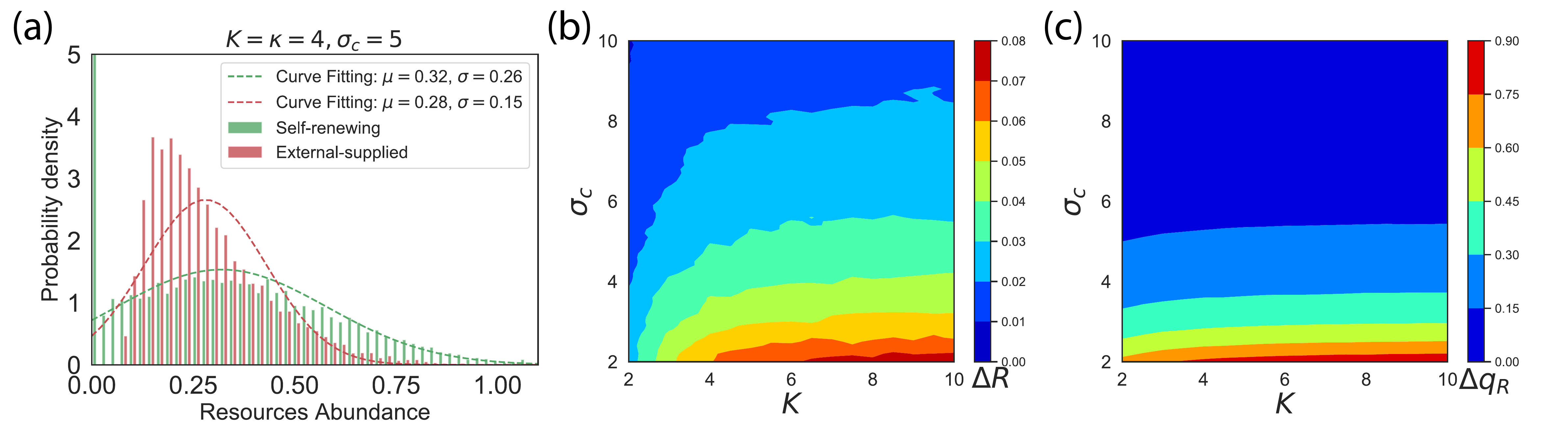}
\caption{(a) Comparison of resource abundance for self-renewing and externally-supplied resource dynamics at $K=\kappa=4$ and $\sigma_c=5$. The dashed lines are the gaussian curve fitting about the abundances with mean $\mu$ and variance $\sigma^2$. (b, c) the difference of the first and second moment of the resource abundance between self-renewing and externally-supplied resource dynamics with the same $K=\kappa$ and  $\sigma_c$, $\Delta R= R^{s}-R^{e}$, $\Delta q_R= q_R^{s}-q_R^{e}$, where the upper label $e$ and $s$ represents  self-renewing and externally-supplied, respectively. All simulations are the same as Fig. 3 in the main text. }
\label{rdistribution}
\end{figure}

\begin{figure}[b]
\centering
\includegraphics[width=0.45\textwidth]{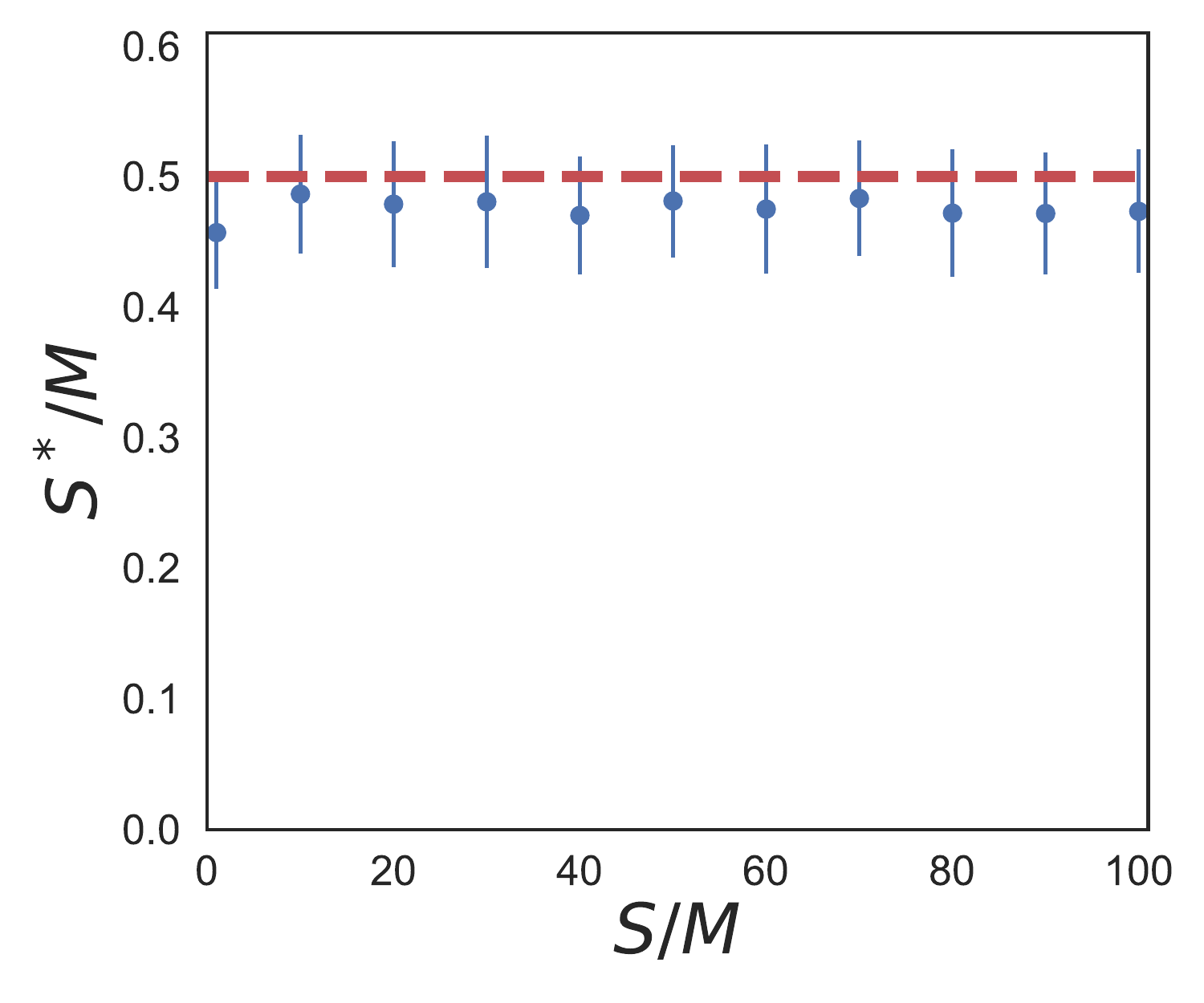}
\caption{The species packing ratio $\frac{S^*}{M}$ at various $S/M$ for externally supplied resource dynamics. Other parameters are the same as Fig. 3 in the main text.}
\label{gamma}
\end{figure}

Eq. (\ref{phiN1}) shows the fraction of surviving species $S^*/S$ is determined by the first moment ($R=\left<R\right>$) and second moment($q_R=\left<R^2\right>$) of the resource abundance.  In Fig. \ref{rdistribution} (a), the simulation shows the two dynamics have similar means but quite different variance for $K=\kappa=4$ and $\sigma_c=5$. And the external-supplied resource dynamics with a larger $q_R$ (sharper distribution) have a smaller fraction of surviving species $S^*/S$. 

Fig. \ref{rdistribution} (b, c) shows the first and second moment differences between self-renewing and externally-supplied resource dynamics,  $\Delta R$ and $\Delta q_R$ are always positive, which means the self-renewing resource dynamics always has larger $R$ and $q_R$ across the whole heat map. And thus, it is generally true that external-supplied dynamics has sharper resource distribution and can explain the lower diversity (in high $\sigma_c$ regime, it looks $\Delta R$ and $\Delta q_R$ are close to zero but considering there is $\sigma_c$ in the dominator of eq. (\ref{phiN1}), a slight difference in $q_R$ can have a huge difference.). However, note that $S^*/S$ (the fraction of species in the regional species pool that survive) is not the same as species packing $S^*/M$ and it cannot explain why the species packing bound is exactly at 0.5.

\end{document}